\pdfoutput=1
\documentclass[prb,twocolumn,showpacs,amsmath,amssymb,floatfix,superscriptaddress]{revtex4}

\usepackage{graphicx}
\usepackage{epsfig}
\usepackage{dcolumn}
\usepackage{bm}

\begin{document}

\title{Charge pumping by magnetization dynamics in magnetic and semi-magnetic tunnel junctions with interfacial Rashba or bulk extrinsic spin-orbit couplings}

\author{Farzad Mahfouzi}
\affiliation{Department of Physics and Astronomy, University of Delaware, Newark, DE 19716-2570, USA}
\author{Jaroslav Fabian}
\affiliation{Institute for Theoretical Physics, University of Regensburg, D-93040 Regensburg, Germany}
\author{Naoto Nagaosa}
\affiliation{Cross-Correlated Materials Research Group (CMRG) and Correlated Electron Research Group (CERG), RIKEN-ASI, Wako, Saitama 351-0198, Japan}
\affiliation{Department of Applied Physics, University of Tokyo, Tokyo 113-8656, Japan}
\author{Branislav K. Nikoli\' c}
\email{bnikolic@udel.edu}
\affiliation{Department of Physics and Astronomy, University
of Delaware, Newark, DE 19716-2570, USA}
\affiliation{Cross-Correlated Materials Research Group (CMRG) and Correlated Electron Research Group (CERG), RIKEN-ASI, Wako, Saitama 351-0198, Japan}

\begin{abstract}
 We develop a time-dependent nonequilibrium Green function (NEGF) approach to the problem of spin pumping by precessing magnetization in one of the ferromagnetic layers within F$|$I$|$F magnetic tunnel junctions (MTJs) or F$|$I$|$N semi-MTJs  in the presence of intrinsic Rashba spin-orbit  coupling (SOC)  at the F$|$I interface or the extrinsic SOC in the bulk of F layers of finite thickness (F-ferromagnet; N-normal metal; I-insulating barrier).  To express the time-averaged pumped charge current, or the corresponding dc voltage signal in open circuits that was measured in recent experiments on semi-MTJs [T. Moriyama {\em et al.}, Phys. Rev. Lett. {\bf 100},  067602  (2008)], we construct a novel solution for the double-time-Fourier-transformed  NEGFs. The two energy arguments of NEGFs in this representation are connected by the Floquet theorem describing multiphoton emission and absorption processes. Within this fully quantum-mechanical treatment of the conduction electrons, we find that: ({\em i}) only in the presence of the interfacial Rashba SOC the non-zero dc pumping voltage in F$|$I$|$N semi-MTJ can emerge at the adiabatic level (i.e., proportional to the microwave frequency); ({\em ii}) a unique signature of this charge pumping phenomenon, where the Rashba SOC resides within the precessing F layer thereby participating in the pumping process, is dc pumping voltage that changes sign as the function of the precession cone angle; ({\em iii}) unlike conventional spin pumping in the absence of SOCs, where one emitted or absorbed microwave photon is sufficient to match the exact solution in the frame rotating with the magnetization, the presence of the Rashba SOC requires to take into account up to ten photons in order to reach the asymptotic value of pumped charge current; ({\em iv}) disorder within F$|$I$|$F MTJs can enhance the dc pumping voltage in the quasiballistic transport regime; ({\em v}) the extrinsic SOC in F$|$I$|$F MTJs causes spin relaxation and eventually the decay of dc pumping voltage which becomes negligible  when the ratio of F layer thickness to the spin-diffusion length is around five.
\end{abstract}

\maketitle
	
\section{Introduction}\label{sec:intro}

The spin pumping by precessing magnetization is a phenomenon where the moving magnetization of a single ferromagnetic layer, driven by microwave radiation under the ferromagnetic resonance conditions (FMR), emits spin current into adjacent normal metal layers.~\cite{Tserkovnyak2005} The emitted spin current is {\em pure}~\cite{Nagaosa2008} in the sense that it is not accompanied by any net charge flux. This effect is termed ``pumping'' because it happens in the absence of any dc bias voltage, and together with closely related adiabatic quantum pumping of charge~\cite{Switkes1999} or spin~\cite{Watson2003} observed in quantum dots, falls in the category of problems where an open quantum system (i.e., a finite many-particle system in contact with particle reservoirs) is exposed to time-dependent periodic externals fields.

Since angular momentum loss carried by emitted pure spin current adds extrinsic contribution to Gilbert damping, spin pumping has initially been observed~\cite{Tserkovnyak2005,Heinrich2007,Gerrits2006} as an increased broadening of FMR spectra upon switching from a single ferromagnet (F) layer to F$|$N multilayers (N-normal metal). Therefore, it is also an essential ingredient to understand~\cite{Taniguchi2008a} critical current switching in experiments~\cite{Chen2006b} on spin-transfer-torque-driven magnetization dynamics. In fact, spin pumping can also be viewed as the Onsager reciprocal phenomenon~\cite{Hals2010} of spin-transfer torque~\cite{Ralph2008} (STT) in which spin current of large enough density injected into a ferromagnetic layer either switches its magnetization from one static configuration to another or generates a dynamical situation with steady-state precessing magnetization.

Recent vigorous experimental efforts have focused on the direct detection of pure spin current generated by coherent macrospin precession in both ferromagnetic metals~\cite{Saitoh2006,Cao2009,Costache2006} and insulators~\cite{Kajiwara2010,Kurebayashi2011} by converting it into dc voltage signal. For example, the experimental techniques employed for this purpose include the inverse spin Hall effect~\cite{Saitoh2006} or the second static F layer as detector within a magnetic tunnel junction (MTJ),~\cite{Cao2009} as well as the precessing F layer itself which can generate  voltage~\cite{Costache2006} at the F$|$N interface by detecting the backflow spin current (due to spin accumulation~\cite{Gerrits2006} driven by pumping into the N layers thinner than the spin-diffusion length).

The origin of the voltage signal of spin pumping in F$|$I$|$F MTJs can be easily understood~\cite{Xiao2008,Tserkovnyak2008,Chen2009} as a two step process: (1) the magnetization dynamics of the left F layer pumps pure spin current across the tunnel barrier (I-insulator); (2) the pumped spin current is then filtered by the analyzing right F layer whose magnetization is static. This generates charge current or, equivalently, dc pumping voltage in an open circuit.~\cite{Xiao2008,Tserkovnyak2008,Chen2009} This voltage is proportional to the frequency $\propto \hbar \omega$ of microwaves due to adiabatic nature of pumping (in the adiabatic regime, formally $\omega \rightarrow 0$ since energy of microwave photons $\hbar \omega$ is smaller than other relevant energy scales in ferromagnetic solids).

On the other hand, surprisingly large voltage signal \mbox{$\propto \hbar \omega$} observed~\cite{Moriyama2008} in F$|$I$|$N semi-MTJs, which do not contain the second analyzing F layer, has remained unexplained in virtually all recent theories~\cite{Xiao2008,Tserkovnyak2008,Chen2009,Chui2008}  of spin pumping in MTJs. Some of these theories~\cite{Tserkovnyak2008,Chen2009} actually predict tiny voltage signal which being non-adiabatic $\propto (\hbar \omega)^2$ is the second-order effect.

Also, there exists several orders of magnitude discrepancy between underestimated voltage signal of spin pumping in F$|$I$|$F MTJs obtained in the scattering theory,~\cite{Xiao2008} experimental data,~\cite{Cao2009} and overestimated voltage signal obtained in the rotating frame approach~\cite{Tserkovnyak2008,Chen2009} or the tunneling Hamiltonian formalism combined with semiclassical modeling of the interplay of spin diffusion and self-consistent screening around interfaces.~\cite{Chui2008} This can be traced to different device setups where scattering approach was applied to MTJs assuming zero~\cite{Xiao2008} spin accumulation in the F layers modeled as semi-infinite leads (justified through assumption that spin-flip rate in F is larger than the tunnel rate), while unrealistically large bulk~\cite{Tserkovnyak2008,Chen2009} or interfacial~\cite{Chui2008} spin accumulation appears in the other two approaches.

The MTJs employed in spin pumping experiments~\cite{Cao2009,Moriyama2008} contain F layers of nanoscale thickness whose short spin-diffusion length~\cite{Brataas2006} can be modeled by sufficiently strong extrinsic spin-orbit scattering. Most importantly, the very recent experiments~\cite{Miron2010} have unveiled a possibility of strong Rashba spin-orbit coupling (SOC) at the F$|$I interface due to structural inversion asymmetry of the multilayered device within which such interfaces reside (e.g., Rashba SOC was detected in N$|$F$|$I multilayers, but not in the N$|$F$|$N ones~\cite{Miron2010}).

However, SOCs have been traditionally neglected in a variety of approaches to spin and charge pumping by magnetization dynamics.~\cite{Tserkovnyak2005,Xiao2008,Chen2009,Zhu2009,Takeuchi2010} The SOCs in the bulk or on the surface of ferromagnetic materials play crucial role in other phenomena, such as the anomalous Hall effect~\cite{Nagaosa2010} or the tunneling anisotropic magnetoresistance~\cite{Gould2004,Moser2007,Wimmer2009,Chantis2007,Gmitra2009,Matos-Abiague2009} (TAMR). Moreover, the study of the interplay between SOC and STT has been recently initiated~\cite{Hals2010,Obata2008,Haney2010,Manchon2011} through theoretical proposals~\cite{Manchon2008} and experimental realizations~\cite{Miron2010,Gambardelle2011} that exploit SOCs for STT-driven magnetization reversal of a single F layer with greatly reduced critical current required when compared to traditional spin valves or MTJs with two non-collinear magnetizations.~\cite{Ralph2008}

In the ``standard model''~\cite{Tserkovnyak2005,Taniguchi2008a} of spin pumping  in magnetic multilayers containing many~\cite{Taniguchi2008a} F and N layers, the magnitude of pumped spin current by F$|$N interfaces is computed quantum-mechanically via the Brouwer scattering formula~\cite{Brouwer1998} which then serves as the boundary condition for the spin-diffusion equation~\cite{Taniguchi2008a} or enters into the so-called circuit theory~\cite{Brataas2006} where device is split into nodes of characteristic size smaller than the spin-diffusion length. Thus, in these frameworks SOCs enter only phenomenologically through finite spin-diffusion length (a spin can be flipped by SOC and magnetic impurities in N or F layers, as well as by magnon scattering in the F layers).

However, these approaches are not applicable to MTJs since the spin accumulation is not well-defined in the insulating barrier.~\cite{Taniguchi2008a} Moreover, even the magnitude of pumped current cannot be obtained from the ``standard model'' formula,~\cite{Tserkovnyak2005} governed by the interfacial spin-mixing conductance~\cite{Brataas2006} $g^{\uparrow\downarrow}$, when {\em strong SOC is present immediately at the interface} which renders $g^{\uparrow\downarrow}$ an ill-defined quantity.

The recent alternative description~\cite{Takeuchi2010} of spin pumping in F$|$N multilayers, based on nonequilibrium Green function (NEGF) expressions for the local spin and charge current densities, has encompassed both the earlier considered~\cite{Silsbee1979} nonlocal diffusion of the spin accumulation at the F$|$N interface generated by magnetization precession and the effective field described by the ``standard model'' (where spin accumulation does not build at the interface since spin-flip relaxation rate is assumed to be sufficiently larger than the spin injection rate). However, this framework~\cite{Takeuchi2010} has treated SOCs only in the N layer away from the precessing F layer in order to analyze how each of these two pumped spin currents can be converted into charge current by the inverse spin Hall effect due to the extrinsic or intrinsic SOCs. That is, in this theory SOCs are not essential for the  discussion of spin pumping effect itself.

We note that other groups have also recently identified the importance of adding SOCs explicitly into the description of spin and charge pumping by magnetization dynamics. For example, a generalized scattering theory of adiabatic charge  pumping by a single precessing F layer within N$|$F$|$N junctions containing SOCs has been formulated in Ref.~\onlinecite{Hals2010}. Also, the pumping of current of magnetic monopoles and the associated (via Amp\`{e}re law) charge current flowing in the plane of the Rashba SO-coupled interface (rather than perpendicular to it as is the focus of our study) between the precessing F layer and a nonmagnetic layer has been predicted in Ref.~\onlinecite{Takeuchi2011}.

\begin{figure}
\includegraphics[scale=0.55,angle=0]{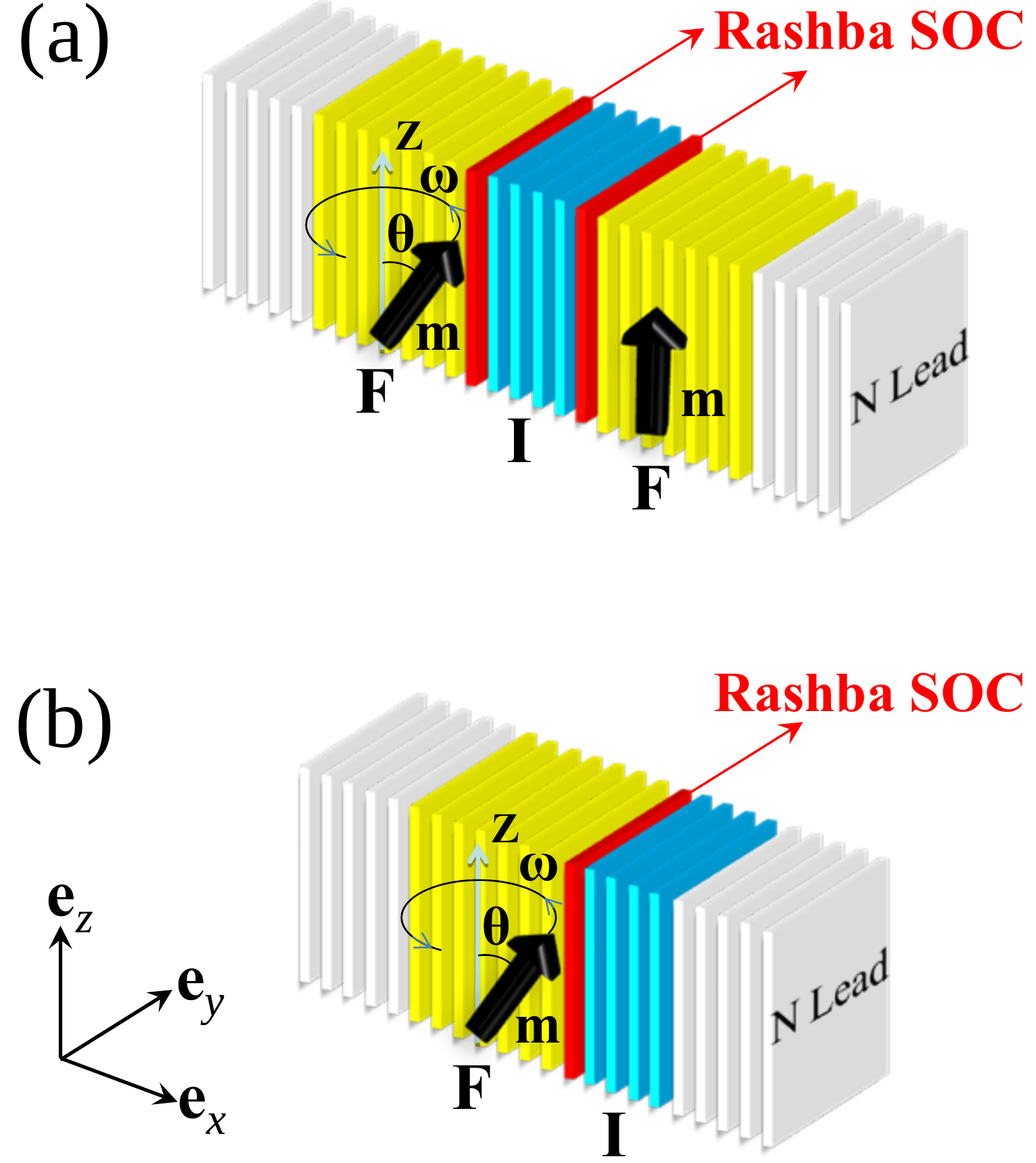}
\caption{(Color online) (a) F$|$I$|$F MTJ and (b) F$|$I$|$N semi-MTJ with precessing magnetization of a single F layer are modeled on a simple cubic finite-size tight-binding lattice attached to semi-infinite ideal (disorder and interaction free) N leads. The thicknesses of the ferromagnetic layers and thin insulating barrier is measured using the number of atomic monolayers $d_F$ and $d_I$, respectively. For example, $d_F=8$ and $d_I=4$ in the illustration, while in the actual calculations we use $d_F=50$ and $d_I=5$ monolayers of cross section $20 \times 20$ lattice sites. The interfacial Rashba SOC due to structural inversion asymmetry of the junction is included in the last monolayer of the F slab that is in direct contact with the tunnel barrier I. The F layers can also include disorder modeled as a random on-site potential and the corresponding extrinsic SOC, while binary alloy disorder in the I layer models AlO$_x$-type tunnel barrier.}
\label{fig:setup}
\end{figure}

Here we develop a NEGF-based quantum transport theory of spin current pumping, its propagation, and conversion into electrically measurable signals in F$|$I$|$F and F$|$I$|$N junctions depicted in Fig.~\ref{fig:setup}. The junctions are described by the microscopic time-dependent (due to precessing magnetization of one of the F layers) Hamiltonian which is tailored to take into account nanoscale thickness of F layers within which we include terms describing disorder and extrinsic SOC in the F layers, as well as possibly strong Rashba SOC at the F$|$I interfaces. Our theory starts from the equations of motion generated by such Hamiltonian for NEGFs which depend on two time variables,  and then finds a computationally efficient solution to such equations which physically describes  processes where a specific number of microwave photons is  absorbed or emitted by propagating electrons in the course of pumping. This solution for time-dependent NEGFs allows us to obtain time-averaged total charge current in the N leads of the junctions shown in Fig.~\ref{fig:setup} or the corresponding dc pumping voltage in the corresponding open circuits. The formulas which we derive for pumped currents are also applicable to any problem where a quantum-mechanical system is exposed to periodic time-dependent external field, independently of its frequency (i.e., including both adiabatic and non-adiabatic regimes) or amplitude (i.e., including both perturbative and non-perturbative regimes), as long as electron-electron, electron-phonon and electron-magnon interactions can be neglected.

The paper is organized as follows. In Sec.~\ref{sec:setup} we discuss how to tune parameters of the microscopic Hamiltonian in order to reproduce properties of MTJs and semi-MTJs employed in experiments,~\cite{Cao2009,Moriyama2008} such as their tunneling magnetoresistance, diffusive nature of transport within the F layers, finite spin-diffusion length in F layers and strong interfacial Rashba SOC. Section~\ref{sec:negf} discusses NEGF equations of motion and how to solve such equations after converting them into algebraic ones via double Fourier transform. In this section we also employ clean F$|$I$|$F  junctions (with semi-infinite or finite thickness F layers) as a test bed to compare our theory to the scattering formulas for pumping in MTJs derived in Ref.~\onlinecite{Xiao2008}, as well as to previously developed~\cite{Wang2003} solution to double-time-Fourier-transformed NEGF equations using continued fractions. In Sec.~\ref{sec:rashba} we discuss properties of time-averaged pumped charge current and the corresponding dc voltage signal in open circuits for clean F$|$I$|$N and F$|$I$|$F junctions as a function of the strength of interfacial Rashba SOC. The effect of disorder in F and I layers, as well as the extrinsic SOC  within F layers, on the dc pumping voltage in F$|$I$|$F MTJs is discussed in Sec.~\ref{sec:disorder}. We conclude in Sec.~\ref{sec:conclusions}. Readers seeking to understand physical effects of SOC on microwave-driven MTJs  may wish to start with Sec.~\ref{sec:setup} and then jump to  Sec.~\ref{sec:rashba} and Sec.~\ref{sec:disorder}, while those interested in theoretical

\section{MTJ device setup and its Hamiltonian}\label{sec:setup}

The MTJ and semi-MTJ we study are illustrated in Fig.~\ref{fig:setup}(a) and (b), respectively. Each atomic monolayer shown in Fig.~\ref{fig:setup} is modeled on the square tight-binding lattice  with single $s$-orbital per site. Since disorder is included as random potential in real space, atomic monolayers are of finite size $20 \times 20$ sites. This can also be viewed as the cross section of a supercell which is repeated periodically in the transverse direction,~\cite{Brataas2006} while its size is sufficient to allow one to compute all quantities at the $\Gamma$ point (i.e., without the need to perform k-point sampling).~\cite{Thygesen2005} We have checked that dc pumping voltage remains constant as one increases cross section size beyond  $20 \times 20$ sites.

The ferromagnetic layers consist of $d_F=50$ such monolayers, so that their thickness is $\simeq 15$ nm (assuming typical lattice spacing \mbox{$a \simeq 3$ \AA}), which closely mimics F layers employed experimentally.~\cite{Cao2009,Moriyama2008} The thickness of the insulating barrier is $d_I=5$ atomic monolayers. The finite-size F$|$I$|$F or F$|$I$|$N multilayer  is connected to macroscopic reservoirs via two semi-infinite ideal (i.e., disorder, spin and charge interaction-free) N leads to form a two-terminal device required for both NEGF and  scattering theory analysis.

The general time-dependent Hamiltonian describing these two devices can be written as
\begin{eqnarray}\label{eq:hamiltonian}
\hat{H}(t) & = & \sum_{{\bf i},ss'} \left(\varepsilon_{\bf i} \delta_{ss'} - \frac{\Delta_{\bf i}}{2} \mathbf{m}_{\bf i}(t) \cdot [\hat{\bm \sigma}]_{ss'} \right) \hat{c}_{{\bf i}s}^{\dagger}\hat{c}_{{\bf i}s'} \nonumber\\
\mbox{}&& - \gamma \sum_{\langle {\bf ii'} \rangle, ss'} \hat{c}_{{\bf i}s}^{\dagger} \hat{c}_{{\bf i}^\prime s'} + i \lambda_\mathrm{ESO} \sum_{{\bf ii'},ss'} \hat{c}_{{\bf i}s}^{\dagger} {\bf L}_{\bf ii'} \cdot [\hat{\bm \sigma}]_{ss'}  \hat{c}_{{\bf i}s'} \nonumber \\
\mbox{}&& + \sum_{\langle {\bf ii'} \rangle, ss'} \hat{c}_{{\bf i}s}^\dag t_{\bf ii'}^{ss'}\hat{c}_{{\bf i'}s'}.
\end{eqnarray}
Its time dependence stems from the unit vector ${\bf m}_{\bf i}(t)$ along the local magnetization direction within the left F layer, which is assumed to be spatially uniform and steadily precessing around the $z$ axis with a constant cone angle. The value of angle $\theta$ is controlled by the input microwave power (typically $\theta \lesssim 20^\circ$ in the recent experiments~\cite{Moriyama2008}). The operators $\hat{c}_{{\bf i}\sigma}^\dag$ ($\hat{c}_{{\bf i}\sigma}$) create (annihilate) electron with spin $\sigma$ at site ${\bf i}=(i_x,i_y,i_z)$, and $\gamma$ is the nearest neighbor hopping which sets the unit energy scale. The coupling of itinerant electrons to collective magnetization dynamics is described through the material-dependent exchange potential $\Delta_{\bf i}$, where $\hat{\bm \sigma}=(\hat{\sigma}_x,\hat{\sigma}_y,\hat{\sigma}_z)$ is the vector of the Pauli matrices and $[\hat{\sigma}]_{ss'}$ denotes the Pauli matrix elements.

The disorder within F layers can be introduced using the uniformly distributed random variable \mbox{$\varepsilon_{\bf i}^F  \in [-W/2,W/2]$} which models isotropic short-range static impurity potential. To account for the properties of amorphous AlO$_x$ tunnel barrier in MTJs and semi-MTJs employed  in the recent spin pumping experiments,~\cite{Cao2009,Moriyama2008} the on-site potential on I monolayers is chosen as $\varepsilon_{\bf i}^I = U_b \pm \delta U_b$ where random fluctuations $\delta U_b$ mimic binary alloy disorder.~\cite{Itoh2003}

The impurity potential  in the F layers also generates extrinsic SOC, as described by the third sum in Eq.~(\ref{eq:hamiltonian}). This can be viewed as the lattice version~\cite{Nikolic2007} of the Thomas term $\lambda ({\bm \sigma} \times {\bf p}) \cdot \nabla V_{\rm dis}$ in the Pauli-Schr\"{o}dinger equation  so that
\begin{eqnarray}\label{eq:eso}
({\bm L}_{\textbf{i} \textbf{i}'})_x=\sum_{\nu,\nu'=\pm 1}\nu \nu' (\varepsilon^F_{\textbf{i} + \nu {\bf e}_y}-\varepsilon^F_{\textbf{i} + \nu'{\bf e}_z})\delta_{\textbf{i},\textbf{i}' + \nu {\bf e}_y + \nu' {\bf e}_z}.
\end{eqnarray}
That is, on the tight-binding lattice the extrinsic SOC acts as additional spin-dependent hopping between both nearest neighbor and next nearest neighbor sites. Here $({\bf e}_x, {\bf e}_y, {\bf e}_z)$ are the unit vectors along the $x,y,z$ axis respectively and $\lambda_{\rm ESO}=\lambda/4a$.

The fourth sum in Eq.~(\ref{eq:hamiltonian}) is the tight-binding representation~\cite{Nikolic2007} of the Rashba SOC written in terms of a generalized nearest neighbor hopping term that acts as $2 \times 2$ Hermitian  matrix in the spin space:
\begin{eqnarray}\label{eq:sohopping}
{\bf t}_{\bf ii'}=\left\{
\begin{array}{cc}
-i\gamma_{\rm RSO} \hat{\sigma}_z & ({\bf i}={\bf i}' + {\bf e}_y)\\
+i\gamma_{\rm RSO} \hat{\sigma}_y &  ({\bf i}={\bf i}'+{\bf e}_z)
\end{array}\right.,
\end{eqnarray}
The continuous version of the Rashba SOC, \mbox{$\alpha_{\rm RSO} (\hat{\bm \sigma}  \times \hat{\bf p}) \cdot {\bf e}_x/\hbar$}, has been traditionally studied in the context of two-dimensional electron gases (2DEGs) within semiconductor heterostructures with structural inversion asymmetry in the growth direction.~\cite{Winkler2003} Nevertheless, several experiments have recently reported evidence of the Rashba SOC-induced splitting of the surface states in both non-magnetic and magnetic metals using angle-resolved photoelectron spectroscopy.~\cite{Bentmann2009} The very recent transport experiments~\cite{Miron2010} have demonstrated Rashba SOC-induced STT of a single thin F layer embedded between two asymmetric interfaces. For example, such effect was observed in Pt$|$Co$|$AlO$_x$ multilayers, but not in the inversion symmetric ones Pt$|$Co$|$Pt. The experiment of Ref.~\onlinecite{Miron2010} has also utilized heavy atoms and surface oxidation to create strong out-of-plane potential gradient in Pt$|$Co$|$AlO$_x$ junctions and enhance the interfacial Rashba SOC.

This motivates the introduction of the Rashba SOC term of strength $\gamma_{\rm RSO}$ into the Hamiltonian in Eq.~(\ref{eq:hamiltonian}), which we set to be  non-zero only on the last monolayer of the precessing F layer that is in the direct contact with the first monolayer of the tunnel barrier, as illustrated in Fig.~\ref{fig:setup}. Note that the exact location of the interfacial Rashba SOC eventually requires fitting the Hamiltonian parameters to first-principles analysis.~\cite{Wimmer2009,Chantis2007,Gmitra2009}

\subsection{How to tune Hamiltonian parameters to reproduce properties of MTJs used in experiments}

\begin{figure}
\includegraphics[scale=0.3,angle=0]{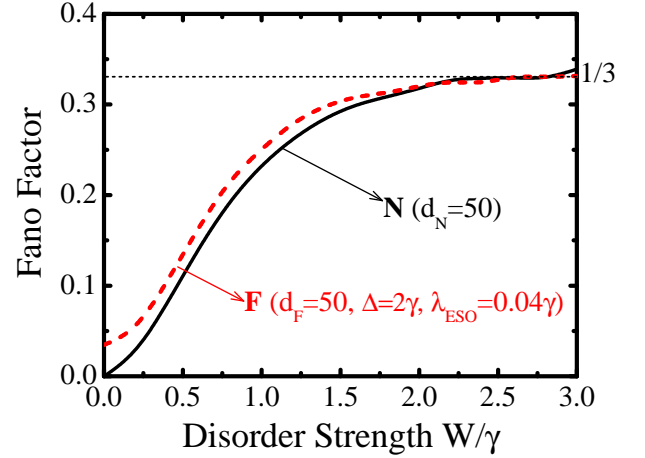}
\caption{(Color online) The Fano factor of the zero-temperature and zero-frequency shot noise  vs. the disorder strength $W$ in transport through N slab with static disorder or F slab with both static disorder and the corresponding extrinsic SOC of strength $\lambda_{\rm ESO}=0.04 \gamma$. Both slabs consists of 50 monolayers (containing $20 \times 20$ atoms per cross section) which are connected to two semi-infinite ideal N leads.}
\label{fig:fano}
\end{figure}

The diffusive transport regime within F or N layers is  defined semiclassically by the requirement that mean free path $\ell$ is smaller then the thickness of such  layers, $\ell \ll d_F$ or $\ell \ll d_N$. In quantum transport calculations, the easiest way to select proper range of values for the disorder strength $W$ which ensures diffusive regime is to compute the Fano factor $F=S/2eI$ of the  shot noise whose zero-temperature and zero-frequency noise power is $S$. For the diffusive metallic wires, $F=1/3$ is universal in the sense of being independent of the impurity distribution, band structure, and shape of the conductor.~\cite{Blanter2000} In Fig.~\ref{fig:fano} we plot the Fano factor with increasing disorder strength $W$ for both conventional disordered N layer attached to two ideal semi-infinite leads and F layer of the same dimensions with both disorder and extrinsic SOC. Using Fig.~\ref{fig:fano}, we select $W=3 \gamma$ to ensure semiclassical diffusive transport regime. 

Note that spin-dependent effects on the shot noise are reveled only when spin-polarized current is injected and the corresponding spin-resolved Fano factors are defined.~\cite{Nikolic2009} Thus, for unpolarized injected charge current, both F and N layers have virtually the same Fano factor in the diffusive regime, while larger Fano factor for the F layer in the quasiballistic regime is due to increased scattering at the N$|$F interfaces because of non-zero $\Delta$.

\subsubsection{Extrinsic SOC in the bulk of F layers}\label{sec:eso}

In both N and F layers, spin-flip scattering will destroy nonequilibrium spin accumulation which is typically accounted~\cite{Brataas2006} through phenomenological spin-diffusion length $L_{\rm sf}$. Over this length scale, an injected spin accumulation loses its polarization so that $L_{\rm sf}$ in ferromagnets defines the magnetoelectrically active region of F layer in contact with N layer. In metallic ferromagnetic materials, $L_{\rm sf}$ ranges~\cite{Brataas2006} from $5$ nm in Ni$_{80}$Fe$_{20}$ (permalloy), which is often employed as precessing F layer in spin pumping experiments,~\cite{Cao2009,Moriyama2008} to 50 nm in Co.

\begin{figure}
\includegraphics[scale=0.3,angle=0]{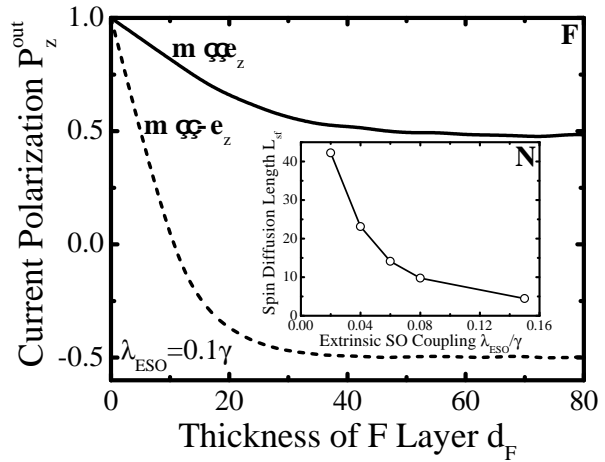}
\caption{The decay of current polarization along the diffusive F layer with static disorder of strength $W=3\gamma$ and the extrinsic SOC of strength $\lambda_{\rm ESO}=0.1 \gamma$. The F layer is attached to two semi-infinite ideal N leads where charge current which is 100 \% spin-polarized along the \mbox{$z$-axis}, ${\bf P}^{\rm in}=(0,0,1)$, is injected from the left N lead and $P_z^{\rm out}$ is computed in the right N lead for F layer of thickness $d_F$. The unit vector of the magnetization in F is either parallel (solid line) or antiparallel (dashed line) to the \mbox{$z$-axis}. The inset shows spin-diffusion length as a function of $\lambda_{\rm ESO}$ when F layer is replaced by diffusive N layer with different strengths of extrinsic SOC, where each value of $L_{\rm sf}$ is extracted by fitting exponentially decaying function to $P_z^{\rm out}$ vs. $d_N$ curves.}
\label{fig:pz}
\end{figure}

To understand how to tune the strength  $\lambda_{\rm ESO}$ of the extrinsic SOC term in Eq.~(\ref{eq:hamiltonian}) in order to generate different experimental situations in F layers of nanoscale thickness, such as  $L_{\rm sf} < d_F$ or $L_{\rm sf} > d_F$ (where spin-flip processes essentially become unimportant), we compute the spin density matrix~\cite{Nikolic2005} of collected transported spins in the right lead $2$ after  fully spin-polarized charge current is injected from the left lead $1$ across the F layer:
\begin{eqnarray} \label{eq:rho_c}
{\bm \rho}^{\rm out} & = &  \frac{e^2/h}{G^{\uparrow \uparrow}_{21} + G^{\downarrow \uparrow}_{21}} \! \sum_{n,m=1}^M \!\!\!
\left( \begin{array}{cc}
     |[{\bf t}_{2 1}^{\uparrow \uparrow}]_{nm}|^2 &  [{\bf t}_{2 1}^{\uparrow \uparrow}]_{nm}
       [{\bf t}_{21}^{\downarrow \uparrow}]_{nm}^*  \\

      [{\bf t}^{\uparrow \uparrow}_{21}]_{nm}^* [{\bf t}^{\downarrow \uparrow}_{21}]_{nm} &
      |[{\bf t}_{21}^{\downarrow\uparrow}]_{nm}|^2
  \end{array} \right) \nonumber \\
\displaystyle  &  = & \frac{1}{2} \left( {\bm 1} + {\bf P}^{\rm out} \cdot \hat{\bm \sigma} \right).
\end{eqnarray}
The elements of the transmission matrix ${\bf t}_{21}^{ss'}$ in this formula  determine the probability $|[{\bf t}_{21}^{\sigma \sigma^\prime}]_{nm}|^2$ for spin-$\sigma^\prime$  electron incident in lead $1$ in the orbital conducting channel $|m\rangle$ to be transmitted to lead $2$ as spin-$\sigma$ electron in channel $|n\rangle$. Therefore, such amplitudes also determine the corresponding spin-resolved conductances,  $G_{2 1}^{ss^\prime}=\frac{e^2}{h} \sum_{n,m=1}^{M} |[{\bf t}_{2 1}^{ss^\prime}]_{nm}|^2$. The spin density matrix makes it possible to extract the detected current polarization vector ${\bf P}^{\rm out}=(P_x^{\rm out},P_y^{\rm out},P_z^{\rm out})$ whose magnitude $P^{\rm out}=|{\bf P}^{\rm out}|$ gives the so-called current polarization measured experimentally.~\cite{Zhu2010} The matrices ${\bf t}_{21}^{\sigma \sigma'}$ and ${\bm \rho}^{\rm out}$ are computed for the two-terminal N$|$F$|$N device where F layer with static magnetization pointing along the \mbox{$z$-axis} and disorder of strength $W=3 \gamma$ is embedded between two semi-infinite N leads [this means that the number of conducting channels $M$ in Eq.~(\ref{eq:rho_c}) is $M=400$ for the selected cross section of $20 \times 20$ sites].

The measured current polarization of permalloy at room temperature ranges from $P=0.32$ to $P=0.5$, depending on the experimental technique employed.~\cite{Zhu2010} Since we find only tiny fluctuations of $P_x, \ P_y \sim 10^{-3}$  in the presence of non-zero $\Delta$ and $\lambda_{\rm ESO}$, we use $P^{\rm out}=|P_z^{\rm out}|$ as the measure of current polarization. We first tune $\Delta=2\gamma$ of the F layer with $\lambda_{\rm ESO}=0$ to obtain $P^{\rm out} \simeq 0.5$ at the Fermi energy $E_F=-3\gamma$. Then we compute the  decay of $P_z^{\rm out}$  with increasing length of the diffusive F layer with non-zero $\lambda_{\rm ESO}$, as shown in Fig.~\ref{fig:pz}. For sufficiently thick F layer, these curves saturate at $|P_z^{\rm out}| \simeq 0.5$. On the other hand, the same calculation for the diffusive N layer with non-zero $\lambda_{\rm ESO}$ gives usual exponentially decaying $P_z^{\rm out}$ vs. $d_N$ curves due to spin diffusion, whose fitting establishes the correspondence between $\lambda_{\rm ESO}$ values used in our study and microscopically determined spin-diffusion length $L_{\rm sf}$. The dependence $L_{\rm sf} \propto 1/\lambda_{\rm ESO}$ shown in the inset in Fig.~\ref{fig:pz}
is expected for the diffusive transport regime.

With $\Delta$ and $E_F$ specified in this fashion to ensure that  current polarization of permalloy slab matches experimentally measured values, we finally tune the height of the potential barrier $U_b=9 \gamma$ in the tunnel barrier I of thickness $d_I=5$ to tune the ``optimistic'' tunneling magnetoresistance (TMR) defined as 
\begin{equation}
{\rm TMR}=\frac{R_{\rm AP}-R_{\rm P}}{R_{\rm P}},  
\end{equation}
to \mbox{TMR=50 \%} for the F$|$I$|$F MTJs employed experimentally.~\cite{Cao2009,Moriyama2008} Here $R_{\rm AP}$ is the resistance with antiparallel configuration of magnetizations in the F layers of thickness $d_F=50$, while $R_{\rm P}$ is the junction resistance when magnetizations are parallel. Since both of these resistances are dominated by the tunnel barrier potential, they are computed for clean junctions.~\cite{Itoh2003} To model AlO$_x$ tunnel barrier, we use binary alloy disorder characterized~\cite{Itoh2003}  by $\delta U_b=0.5 \gamma$.

\subsubsection{TAMR and spin dephasing in perpendicular transport  through interfaces with the Rashba SOC}\label{sec:tamr}

\begin{figure}
\includegraphics[scale=0.35,angle=0]{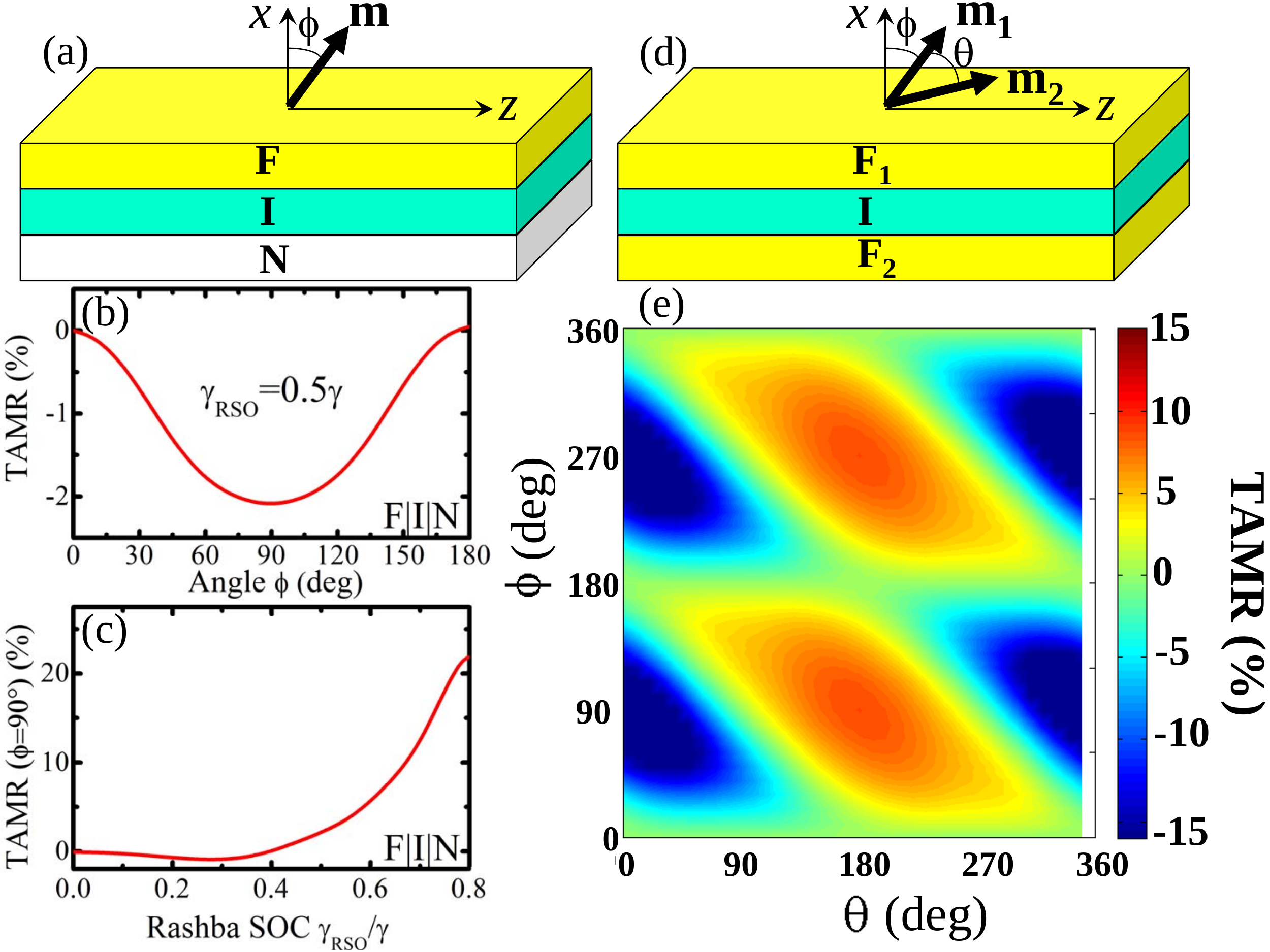}
\caption{(Color online) (a) The setup for the measurement of the out-of-plane TAMR in F$|$I$|$N semi-MTJ, defined by Eq.~(\ref{eq:tamr}), as a function of the angle $\phi$ between the static magnetization of the F layer and the transport direction (the \mbox{$x$-axis}). In panel (b), the Rashba SOC at the F monolayer in contact with the tunnel barrier I is fixed at $\gamma_{\rm RSO}=0.5\gamma$, while panel (c) shows $\mathrm{TAMR}\,(\phi=90^\circ)$ for different values of $\gamma_{\rm RSO}$. (d) The setup for the measurement of the out-of-plane TAMR in F$|$I$|$F MTJ, defined by Eq.~(\ref{eq:tamrmtj}), as a function of the magnetization orientation in each of the two F layers with respect to the transport direction. The TAMR depends on the absolute magnetization directions ${\bf m}_1$ and ${\bf m}_2$. In panel (e),  the Rashba SOC of strength $\gamma_{\rm RSO}=0.5\gamma$ is present at both F monolayers in contact with the tunnel barrier I. The F layers in both semi-MTJ and MTJ have finite thickness $d_F=50$.}
\label{fig:tamr}
\end{figure}

To understand the correspondence between the strength of the Rashba SOC measured by the spin-dependent hopping parameter $\gamma_{\rm RSO}$ in Eq.~(\ref{eq:sohopping}) and the values encountered in experimental devices,~\cite{Miron2010} we compute the so-called out-of-plane TAMR coefficient for F$|$I$|$N semi-MTJ which is defined as~\cite{Matos-Abiague2009}
\begin{equation}\label{eq:tamr}
\mathrm{TAMR}\,(\phi)=\frac{R(\phi) - R(0)}{R(0)},
\end{equation}
for the device setup illustrated in Fig.~\ref{fig:tamr}(a). In dc transport measurements of TAMR, the magnetization direction in F layer provides a control knob orienting the spin, while the magnetic anisotropy is determined by the interface symmetry rather than by the symmetry of the bulk materials. Here $R(0)$ is the resistance of semi-MTJ when static magnetization of its F layer is parallel to the \mbox{$x$-axis} as the direction of transport in Fig.~\ref{fig:tamr}(a), and $R(\phi)$ is the junction resistance when magnetization is rotated by an angle $\phi$ with respect to the \mbox{$x$-axis} within the $xz$-plane. 

\begin{figure}
\includegraphics[scale=0.3,angle=0]{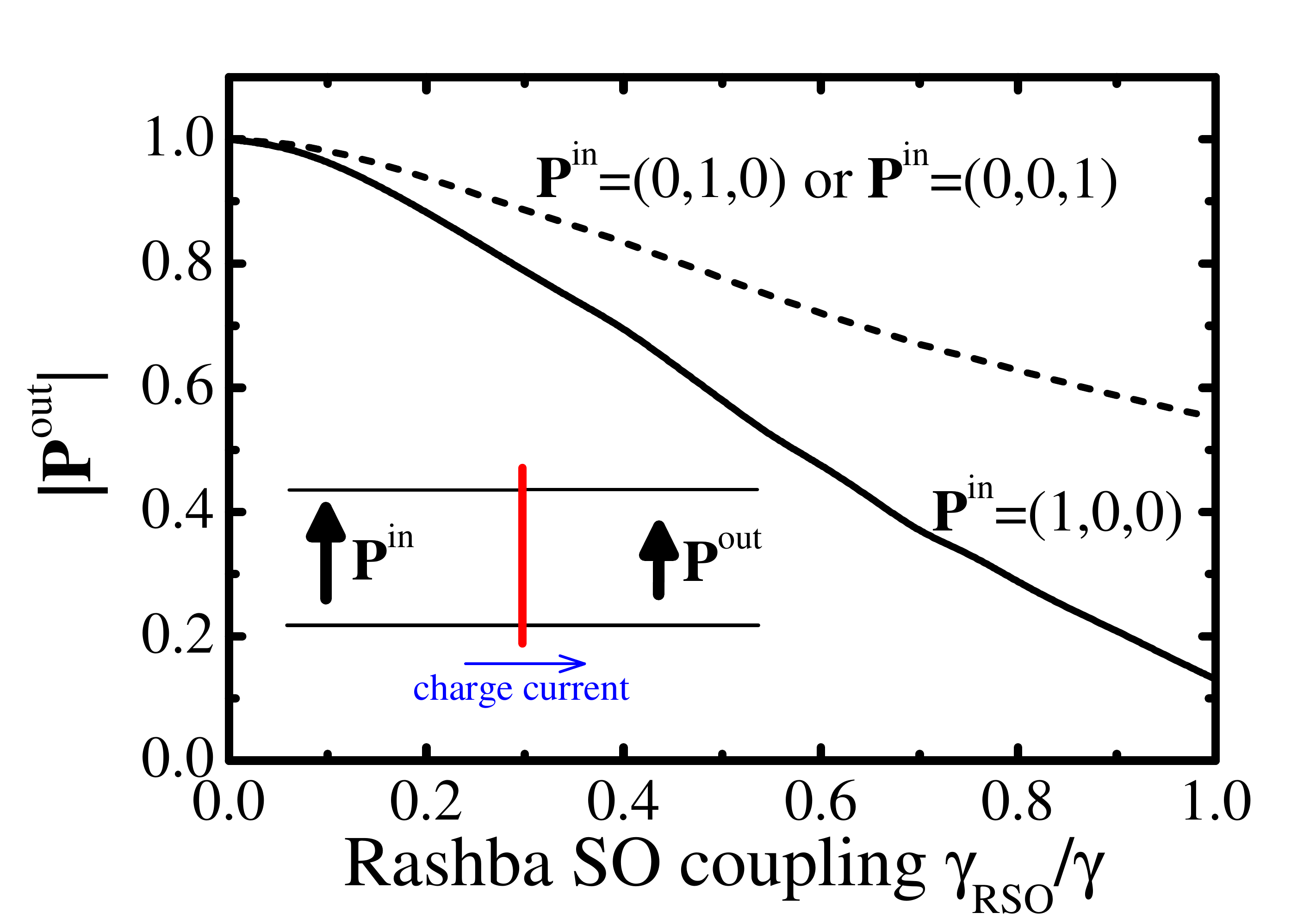}
\caption{(Color online) The magnitude $|{\bf P}^{\rm out}|$ of the spin-polarization vector of outgoing charge current in the right N lead after fully $|{\bf P}^{\rm in}|=1$ spin-polarized current is injected from the left N lead traversing a monolayer with the Rashba SOC of strength $\gamma_{\rm RSO}$. The spin-polarization vector ${\bf P}^{\rm in}$ can point along three different axes of the coordinate system in Fig.~\ref{fig:setup}, where ${\bf P}^{\rm in}=(0,1,0)$ and ${\bf P}^{\rm in}=(0,0,1)$ are parallel to the Rashba monolayer while  ${\bf P}^{\rm in}=(1,0,0)$ is orthogonal to the Rashba monolayer.  The direction of ${\bf P}^{\rm out}$ remains collinear with ${\bf P}^{\rm in}$, as illustrated in the inset.}
\label{fig:pout}
\end{figure}

Figure~\ref{fig:tamr}(b) shows $\mathrm{TAMR}(\phi)$ at fixed Rashba SOC, while the maximum $\mathrm{TAMR}(\phi=90^\circ)$ vs. the strength of Rashba SOC is plotted in Fig.~\ref{fig:tamr}(c). Compared to the weak Rashba SOC in 2DEGs where typically $\gamma_{\rm RSO} \simeq 0.01 \gamma$, the interfacial Rashba SOC in semi-MTJs has to be rater strong (as achieved in the recent experiments~\cite{Miron2010}) in order to generate observable TAMR. Since the interfacial SOI is linear in momentum, TAMR vanishes at the first order in  $\gamma_{\rm RSO}$ after averaging over the Fermi sphere. However the ferromagnet contains local exchange field and a net transfer of angular momentum occurs at the second order, so that TAMR $\propto \gamma_{\rm RSO}^2$. This region of small TAMR occurs for $\gamma_{\rm RSO} \lesssim 0.4 \gamma$ in Fig.~\ref{fig:tamr}(c), beyond which higher order processes start to play the role and TAMR increases faster with increasing $\gamma_{\rm RSO}$.

For F$|$I$|$F junctions, one can define the out-of-plane TAMR coefficient as~\cite{Matos-Abiague2009}
\begin{equation}\label{eq:tamrmtj}
\mathrm{TAMR}\,(\theta,\phi)=\frac{R(\theta,\phi) - R(\theta,0)}{R(\theta,0)},
\end{equation}
where the meaning of angles $\theta$ and $\phi$ is explained in Fig.~\ref{fig:tamr}(d). Since TAMR coefficient for F$|$I$|$N has only one angle argument, there is no ambiguity in using the same TAMR notation for both cases. The out-of-plane TAMR for F$|$I$|$F MTJ is shown in Fig.~\ref{fig:tamr}(e).

Unlike amply studied lateral spin transport in 2DEGs~\cite{Nikolic2005,Nikolic2009} or interfaces~\cite{Miron2010} under the influence of the Rashba SOC,~\cite{Nikolic2005} very little is known about the effect of such interfaces on spin transport perpendicular to the plane, as illustrated by the measurement geometry in the inset of Fig.~\ref{fig:pout}. We clarify their effect by using Eq.~(\ref{eq:rho_c}) to obtain the spin-polarization vector ${\bf P}^{\rm out}$ of the current in the right N lead after 100\% spin-polarized charge current with $|{\bf P}^{\rm in}|=1$ is injected from the left N lead. The result in Fig.~\ref{fig:pout} shows spin dephasing, where the outgoing spin polarization vector ${\bf P}^{\rm out}$ remains in the same direction as ${\bf P}^{\rm in}$, but with reduced magnitude $|{\bf P}^{\rm out}|<1$. The degree of dephasing depends on the direction (perpendicular or parallel) of the initial spin polarization with respect to the Rashba interface.

\section{NEGF approach to pumping by precessing magnetization}\label{sec:negf}

Theoretical studies of  quantum charge pumping in noninteracting phase-coherent systems have been conducted using a variety of approaches. In the adiabatic regime, Brouwer scattering formula~\cite{Brouwer1998} is often used as an elegant geometrical description of the charge pumped per cycle in terms of the instantaneous scattering matrices of the system. The adiabatic regime occurs when time-dependence of the driving field parameters is slow in comparison to the characteristic time scales of the system, such as the electron dwell time, so that electrons traverse the device as if the external potential landscape if frozen in time.  Approaches beyond adiabatic regime include Floquet scattering theory,~\cite{Moskalets2002} iterative solutions of time-dependent states~\cite{Entin-Wohlman2002} and variations of the NEGF formalism.~\cite{Wang2003,Arrachea2006} Moreover, the generality of the time-dependent NEGF framework~\cite{Jauho1994}  makes it a usual choice in the studies of pumping in the presence of strong Coulomb interactions.~\cite{Fioretto2008}

Among these approaches, Brouwer scattering formula~\cite{Tserkovnyak2005} and NEGF formalism~\cite{Chen2009} have been employed to describe experiments on spin pumping by moving magnetization in magnetic multilayers. Unlike quantum charge pumping, the spin pumping in magnetic multilayers is robust and ubiquitous effect at room temperature. Nevertheless, the match between Brouwer scattering formula~\cite{Tserkovnyak2005} and experiments on F$|$N multilayers is excellent due to pumped spin current being determined by the processes at the F$|$N interface. The scattering theory expresses pumped current by a remarkably simple formula
\begin{equation}\label{eq:rmp2005}
I^S {\bf P} =\frac{\hbar}{4\pi} {\rm Re}\, g^{\uparrow\downarrow} \, {\bf m} \times \frac{d{\bf m}}{dt},
\end{equation}
whose dc component is given by
\begin{equation}\label{eq:isin}
I^{S_z} = \frac{\hbar \omega}{4\pi}  {\rm Re}\, g^{\uparrow\downarrow} \sin^2 \theta.
\end{equation}
However, the derivation leading to this formula, as well as the very definition of the spin-mixing conductance~\cite{Brataas2006} $g^{\uparrow\downarrow}$ (where ${\rm Re} \, g^{\uparrow\downarrow}$ is its real part) of the F$|$N interface, assumes absence of any spin-flips.~\cite{Tserkovnyak2005}

Naively, one could numerically evaluate the Brouwer scattering formula for the whole device without introducing ill-defined $g^{\uparrow\downarrow}$ in the presence of interfacial SOCs. However, SOC renders all components of pumped spin or charge current
time-dependent so that one has to compute the scattering matrix {\em at all times} within one period $\tau=2\pi/\omega$ of the pumping cycle
and then find the time-averaged value of pumped currents.~\cite{Hals2010,Ohe2008} This is prohibitively expensive for 3D system composed of large number of atomic orbitals [such as the device in Fig.~\ref{fig:setup}(a) whose Hamiltonian matrix is of the size $84000 \times 84000$],  especially in the presence of disorder where additional averaging over impurity configurations is required. Although this could be achieved for smaller device sizes, we find that the maximum value of pumped current oscillating in time is orders of magnitude larger than its average value over one period in the case of MTJs which prevents the estimate of experimentally relevant time-averaged values from the numerical data.

The rotating frame approach,~\cite{Tserkovnyak2008,Chen2009} where pumping due to precessing magnetization is mapped onto a dc transport within a four-terminal device whose currents can be computed~\cite{Chen2009} using NEGFs, is also inapplicable in the presence of SOCs or other spin-flip mechanisms. This is due to the fact that the same unitary transformation (discussed in Sec.~\ref{sec:comparerf}) which maps time-dependent Zeeman term in Eq.~(\ref{eq:hamiltonian}) to the one frozen at $t=0$, generates new time-dependent SOC terms in the rotating frame.

The time-dependent NEGF formalism~\cite{Jauho1994,Haug2007} makes it possible to obtain directly the time-averaged current. The intricacy in solving its equations stems from the fact that nonequilibrium problems are not time-translation invariant, so one has to work with equations for most general case of NEGFs which depend on two time variables. There are two independent GFs that need to be determined: (1) the retarded GF
\begin{equation}\label{eq:retarded}
G^{r}_{{\bf ii'},{ss'}}(t,t') = - i \Theta(t-t') \langle \{\hat{c}_{{\bf i}s}(t) , \hat{c}^\dagger_{{\bf i'}s'}(t')\}\rangle,
\end{equation}
describes the density of available quantum-mechanical states;  and (2) the lesser GF
\begin{equation}\label{eq:lesser}
G^{<}_{{\bf ii'},{ss'}}(t,t') = i \langle \hat{c}^\dagger_{{\bf i'}s'}(t') \hat{c}_{{\bf i}s}(t) \rangle,
\end{equation}
determines how electrons occupy those quantum states. Here $\langle \ldots \rangle$ denotes the nonequilibrium statistical average,~\cite{Haug2007} and we use  $\hbar=1$ to simplify notation in this Section. Both GFs can be extracted from the contour-ordered GF defined for any two-time values that lie on the Keldysh contour~\cite{Haug2007} (consisting of a two-way path that begins at $-\infty$ time, draws forward to $+\infty$, and then backward to $-\infty$).

The retarded GF is governed by the following equation of motion~\cite{Jauho1994}
\begin{widetext}
\begin{equation}\label{eq:dyson}
i\frac{\partial}{\partial t} \textbf{G}^r(t,t') - \textbf{H}(t) \textbf{G}^r(t,t') -\int\limits_{-\infty}^{+\infty} dt'' \, \bm{\Sigma}^r(t - t'')\textbf{G}^r(t'',t')=\delta(t-t'),
\end{equation}
where we use ${\bf G}^r$, ${\bf H}$(t), and ${\bm \Sigma}^r(t)$ notation to emphasize that these are matrices whose indices represent space and spin degrees of freedom. In noninteracting systems, the retarded self-energy ${\bm \Sigma}^r(t) = \sum_p {\bm \Sigma}_p^r(t)$ is simply the sum of self-energies ${\bm \Sigma}^r_p(t)$ due to leads $p$ attached to the sample.

The lesser GF satisfies the Keldysh integral equation
\begin{eqnarray}\label{eq:keldysh}
\textbf{G}^<(t,t')=\int\limits_{-\infty}^{+\infty} dt_1 \! \int\limits_{-\infty}^{+\infty} dt_2 \, \textbf{G}^r (t,t_1) {\bf{\Sigma}}^<(t_1,t_2)\textbf{G}^a(t_2,t'),
\end{eqnarray}
where the advanced GF is related to the retarded one through ${\bf G}^r(t,t')=[{\bf G}^a(t',t)]^\dagger$. In the case of noninteracting systems, the lesser self-energy ${\bm \Sigma}^<(t) = \sum_p {\bm \Sigma}_p^<(t)$ is only due to attached leads, which can be obtained from the retarded ones using ${\bm \Sigma}^<_p = if_p{\bm \Gamma}_p(t)$. Here $f_p$ is the Fermi function of the macroscopic reservoir to which the lead $p$ is attached at infinity and ${\bm \Gamma}_p(t)=i({\bm \Sigma}_p^r - [{\bm \Sigma}_p^r(t)]^\dagger)$.

The Heisenberg equation for the charge $\hat{Q}=e\sum_{{\bf i}\in p,ss'} \hat{c}_{{\bf i}s}^{\dagger}\hat{c}_{{\bf i}s'}$ or spin density $\hat{S}_\alpha=1/2 \sum_{{\bf i} \in p,ss'} \hat{c}_{{\bf i}s}^{\dagger} [\sigma_\alpha]_{ss'} \hat{c}_{{\bf i}s'}$ operators of electrons in lead $p$ then yields expressions for time-dependent total charge current~\cite{Jauho1994}
\begin{equation}
I_p (t) = \left \langle \frac{d \hat{Q}}{dt} \right \rangle = - 2e \, {\rm Re} \int\limits_{-\infty}^{+\infty} dt' {\rm Tr}\, [{\bf G}^r(t,t') {\bm \Sigma}_p^<(t' - t) + {\bf G}^<(t,t') {\bm \Sigma}^a_p(t'-t)],
\end{equation}
or spin current
\begin{equation}
I_p^{S_\alpha} (t) = \left \langle \frac{d \hat{S}}{dt} \right \rangle = -  {\rm Re} \int\limits_{-\infty}^{+\infty} dt' {\rm Tr}\, \{ \sigma_\alpha [{\bf G}^r(t,t') {\bm \Sigma}_p^<(t' - t) + {\bf G}^<(t,t') {\bm \Sigma}^a_p(t'-t)] \},
\end{equation}
\end{widetext}
in lead $p$. Here ${\bm \Sigma}^a_p(t) = [{\bm \Sigma}^r_p(-t)]^\dagger$. The integration of these expressions, $\lim_{T \rightarrow \infty} \frac{1}{T} \int_{-T/2}^{T/2} (\ldots)$, yields the corresponding time-averaged quantities $I_p$ and $I_p^{S_\alpha}$.

In stationary problems ${\bf G}^r(t,t')$ and ${\bf G}^<(t,t')$ depend on the time-difference $t-t'$, which allows to Fourier transform them into functions of a single energy argument and reduce the set of coupled integral and integro-differential equations to a set of algebraic equations. On the other hand, when the device Hamiltonian depends on time explicitly, one has to work with both times. Since directly solving equations Eq.~(\ref{eq:dyson}) and (\ref{eq:keldysh}) is cumbersome, it is advantageous to switch to a more convenient representation. The typical choices used for problems containing periodic time-dependent fields are: ({\em i}) the double-time Fourier transform~\cite{Wang2003,Li2007a,Hattori2008}
\begin{equation}\label{eq:doubleft}
{\bf G}^{r,<}(t,t')=\int\limits_{-\infty}^{+\infty} \frac{dE}{2\pi} \! \int\limits_{-\infty}^{+\infty}  \frac{E'}{2\pi} \, e^{-i E t + i E' t'} {\bf G}^{r,<}(E,E');
\end{equation}
({\em ii}) the single Fourier transform~\cite{Arrachea2007} in the time difference $t-t'$
\begin{equation}\label{eq:singleft}
{\bf G}^{r,<}(t,t') = \int\limits_{-\infty}^{+\infty} \frac{dE}{2\pi} \, e^{-iE(t-t')}{\bf G}^{r,<}(t,E);
\end{equation}
and ({\em iii}) the so-called Floquet matrix form ${\bf G}^{r,<}_{mn}(\omega)$ defined by~\cite{Tsuji2008}
\begin{eqnarray}\label{eq:floquetgf}
{\bf G}_n^{r,<}(E) &  = & \int\limits_{-\infty}^{+\infty} dt_{\rm rel} \frac{1}{\tau} \int_{-\tau/2}^{\tau/2} dt_{\rm av} e^{i E  t_{\rm rel} + i n \Omega t_{\rm av}} {\bf G}^{r,<}(t,t'), \nonumber \\
{\bf G}_{mn}^{r,<}(\omega) & = & {\bf G}^{r,<}_{m-n} \left(E + \frac{m+n}{2} \Omega \right).
\end{eqnarray}
The expressions in Eq.~(\ref{eq:floquetgf}), where \mbox{$t_{\rm rel}=t-t'$} and \mbox{$t_{\rm av}=(t+t')/2$}, exploit the periodicity condition
${\bf G}(t+\tau,t'+\tau)={\bf G}(t,t')$ and are, therefore, GF counterpart of the Floquet matrix representation for periodically
time-dependent Hamiltonian  $\hat{H}(t+\tau)=\hat{H}(t)$ and its eigenstates (with the ``Brillouin zone'' of energies being $-\Omega/2 < E \le \Omega/2)$.

\subsection{Exact multiphoton solution to double-time-Fourier-transformed NEGFs}\label{sec:onephoton}

Here we adopt the double-time Fourier transform in Eq.~(\ref{eq:doubleft}), which has been used frequently to solve NEGF equations for non-adiabatic charge pumping~\cite{Wang2003} or spin pumping~\cite{Li2007a,Hattori2008} from the 2DEG with the Rashba SOC driven by  time-periodic external fields. Due to the Floquet theorem, the double-time-Fourier-transformed retarded GF ${\bf G}^r(E,E')$ must take the form
\begin{equation}\label{eq:multiphotongf}
{\bf G}^r(E,E')={\bf G}^r(E,E+n\omega)={\bf G}^r_n(E).
\end{equation}
The coupling of energies $E$ and $E+ n\omega$ ($n$ is integer) indicates how multiphoton exchange processes contribute toward the pumped current.

The double-time-Fourier-transformed Eq.~(\ref{eq:dyson}) is given by:
\begin{eqnarray}\label{eq:ftdyson}
&& E \textbf{G}^r(E,E+n\omega) - \int\limits_{-\infty}^{+\infty}  \frac{dE'}{2\pi} \, \textbf{H}(E-E')\textbf{G}^r(E',E+n\omega) \nonumber \\
\mbox{}&&-{\bf{\Sigma}}^r(E)\textbf{G}^r(E,E+n\omega)=2 \pi \delta(n\omega)
\end{eqnarray}
The Fourier transform of the Hamiltonian
\begin{equation}\label{eq:fthamiltonian}
\textbf{H}(E)=2\pi[\textbf{H}_0 \delta(E) + \textbf{V} \delta(E+\omega) + \textbf{V}^{\dagger}\delta(E-\omega)],
\end{equation}
consists of the first term due to the time-independent part ${\bf H}_0$ of Eq.~(\ref{eq:hamiltonian}), while the other two terms are Fourier-transformed harmonic time-dependent part cast in the form ${\bf H}'(t) = {\bf V} e^{i \omega t} + {\bf V}^\dagger e^{-i \omega t}$. Here ${\bf V}$ is the matrix representation of the operator
\begin{equation}\label{eq:potential}
\hat{V} =   - \frac{1}{4} \sum_{{\bf i},ss'} \Delta_{\bf i}   \sin \theta \left( [\sigma_x]_{ss'} - i[\sigma_y]_{ss'} \right) \hat{c}_{{\bf i}s}^{\dagger}\hat{c}_{{\bf i}s'},
\end{equation}
extracted from Eq.~(\ref{eq:hamiltonian}) as the term carrying the periodic time-dependence.

By substituting Eq.~(\ref{eq:fthamiltonian}) into Eq.~(\ref{eq:ftdyson})  we arrive at the following equation:
\begin{eqnarray}\label{eq:fock}
[E \check{\textbf{1}} + \check{\bm \Omega} - \check{\textbf{H}} - \check{\bm \Sigma}^r (E + \check{\bm \Omega})] \check{\textbf{G}}^r (E) = \hat{\textbf{1}}.
\end{eqnarray}
To simplify the notation, we use
\begin{equation}\label{eq:hmatrix}
\check{\bf H}=\left(
\begin{array}{ccccccc}
\ddots & \cdots & \cdots & \cdots & \cdots & \cdots & \cdots \\
\cdots & \ddots & 0 & 0 &  0 & \cdots & \cdots \\
\cdots & \cdots & {\bf H}_0 & {\bf V} &  0 & \cdots & \cdots \\
\cdots & \cdots & {\bf V}^\dagger  & {\bf H}_0 & {\bf V} & \cdots & \cdots \\
\cdots & \cdots & 0 & {\bf V}^\dagger & {\bf H}_0 & \cdots & \cdots \\
\cdots & \cdots & 0 & 0 & 0 & \ddots &  \cdots \\
\cdots & \cdots & \cdots & \cdots & \cdots & \cdots & \ddots \\
\end{array}\right),
\end{equation}
and
\begin{equation}\label{eq:omegamatrix}
\check{\bm \Omega}=\left(
\begin{array}{ccccccc}
\ddots & \cdots & \cdots & \cdots & \cdots & \cdots & \cdots \\
\cdots & -2 \omega {\bf 1} & 0 & 0 &  0 & \cdots & \cdots \\
\cdots & \cdots & -\omega {\bf 1} & 0 &  0 & \cdots & \cdots \\
\cdots & \cdots & 0  & 0 & 0 & \cdots & \cdots \\
\cdots & \cdots & 0 & 0 & \omega {\bf 1} & \cdots & \cdots \\
\cdots & \cdots & 0 & 0 & 0 & 2 \omega {\bf 1} &  \cdots \\
\cdots & \cdots & \cdots & \cdots & \cdots & \cdots & \ddots \\
\end{array}\right).
\end{equation}
Here symbol $\check{\bf A}$ is used to denote a matrix which acts in the Hilbert space ${\mathcal H}_{\rm el} \otimes {\mathcal H}_{\rm ph}$, where the dimension of the Hilbert space of photons ${\mathcal H}_{\rm ph}$ is infinite.  The unit matrix in the Hilbert space of a single electron ${\mathcal H}_{\rm el}$ is ${\bf 1}$, and the unit matrix in ${\mathcal H}_{\rm el} \otimes {\mathcal H}_{\rm ph}$ is denoted by $\check{\mathbf 1}$.

Since higher order multiphoton processes yield progressively smaller contribution to the pumped current, we restrict the dimension  of ${\mathcal H}_{\rm ph}$ by considering up to $N_{\rm ph}$ photons. In this case, the dimensionality of ${\mathcal H}_{\rm ph}$ is $2N_{\rm ph}+1$ since one can have processes with no photon exchange $n=0$ or the maximum of $n=N_{\rm ph}$ photons is absorbed or emitted. This means that if we keep only $N_{\rm ph}=1$ processes, $\check{\bf G}(E)$ is a matrix of the dimension $l_{\rm sites} \times 2_s \times 3_{\rm ph}$ where $l_{\rm sites}$ is the number of lattice sites within the central region of devices in Fig.~\ref{fig:setup}, $2_s$ takes into account the spin degrees of freedom, and $3_{\rm ph}$ is due to single (or no) photon exchange in the course of pumping.

The Keldysh equation~(\ref{eq:keldysh}) in this representation is written as:
\begin{eqnarray}
\check{\textbf{G}}^<(E)=\check{\textbf{G}}^r(E)\check{\bm \Sigma}^< (E + \check{\bm \Omega}) \check{\textbf{G}}^a(E).
\end{eqnarray}
The knowledge of $\check{\bf G}^<(E)$ makes it possible to obtain the time-averaged total pumped charge current in lead $p$ in the absence of any dc bias voltage
\begin{eqnarray}\label{eq:central}
 I_p = \frac{e}{2N_{\rm ph}} \int\limits_{-\infty}^{+\infty} dE \, {\rm Tr} \, [\check{\bm \Gamma}_p \check{\textbf{f}} \check{\textbf{G}}^r \check{\bm \Gamma} \ \check{\textbf{G}}^a - \check{\bm \Gamma}_p \check{\textbf{G}}^r \check{\bm \Gamma} \check{\textbf{f}} \check{\textbf{G}}^a],
\end{eqnarray}
where $\check{\bm \Gamma} = \sum_p \check{\bm \Gamma}_p$ and $\check{\textbf{f}}=f(E + \check{\bm \Omega})$. Since the trace in the integrand, ${\rm Tr} \equiv {\rm Tr}_{\rm el} {\rm Tr}_{\rm ph}$, is summing over contributions from different photon exchange processes, the denominator includes $2N_{\rm ph}$  to avoid double counting. Note that the part of the trace operating in ${\mathcal H}_{\rm ph}$ space ensures the current conservation in our solution to NEGF equations. The analogous formula for the pumped spin current into lead $p$ in the absence of any dc bias voltage is given by
\begin{eqnarray}\label{eq:centralspin}
I_p^{S_\alpha} = \frac{1}{4N_{\rm ph}} \int\limits_{-\infty}^{+\infty}
dE \, {\rm Tr} \, [\hat{\sigma}_\alpha (\check{\bm \Gamma}_p \check{\textbf{f}}
\check{\textbf{G}}^r \check{\bm \Gamma} \ \check{\textbf{G}}^a
- \check{\bm \Gamma}_p \check{\textbf{G}}^r \check{\bm \Gamma}
\check{\textbf{f}} \check{\textbf{G}}^a)].
\end{eqnarray}
Equations~(\ref{eq:central}) and ~(\ref{eq:centralspin}) are the {\em central outcome} of our formalism and can be applied  to arbitrary charge or spin pumping problem.

For the specific problem of pumping by precessing magnetization driven by microwaves, we take into account that $\hbar \omega \ll E_F$ and  simplify Eq.~(\ref{eq:central}) accordingly  by expanding the Fermi function
\begin{equation}
\check{\bf f} = f(E_F)\check{\bf 1} - \check{\bm \Omega} \partial f/\partial E.
\end{equation}
This leads to the following {\em adiabatic} expression for pumped charge current:
\begin{equation} \label{eq:centraladiabatic}
I_p = \frac{e}{2 N_{\rm ph} } {\rm Tr}\, \{ \check{\bf{\Gamma}}_p(E_F) [\check{\bf{\Omega}},\check{\textbf{G}}^r(E_F) \check{\bf{\Gamma}}(E_F)] \check{\textbf{G}}^a(E_F)\}
\end{equation}
assuming zero temperature. The commutator
\begin{equation}
[\check{\bf{\Omega}},\check{\textbf{G}}^r(E_F) \check {\bf{\Gamma}}(E_F)] = \check{\bf{\Omega}}\check{\textbf{G}}^r(E_F)\hat{\bf{\Gamma}}(E_F) - \check{\textbf{G}}^r(E_F) \check{\bf{\Gamma}}(E_F) \check{\bf{\Omega}}
\end{equation}
allows us to make the notation more compact.

The time-averaged value of the pumped charge current is translated into the dc pumping voltage in an open circuit via
\begin{equation}\label{eq:vpump}
V_{\rm pump}=\frac{I_p}{G(\theta)},
\end{equation}
which is the quantity measured in the recent experiments.~\cite{Cao2009,Moriyama2008} Here $G(\theta)$ is the conductance of F$|$I$|$F or F$|$I$|$N  junctions computed by tilting the {\em static} magnetization of the first F layer by an angle $\theta$ away from the $z$-axis and by applying the linear-response bias voltage between the N leads attached to the junction.

\subsection{Comparison with the rotating frame approach}\label{sec:comparerf}

In the absence of interactions causing spin-flips, such as the SOC, it is possible to convert the complicated time-dependent transport problem posed by the presence of precessing magnetization in the devices in Fig.~\ref{fig:setup} into the time-independent one by performing the unitary~\cite{Tserkovnyak2008,Chen2009,Hattori2007,Zhang2003a} transformation of the Hamiltonian Eq.~(\ref{eq:hamiltonian})
\begin{equation}\label{eq:hrot}
\hat{H}_{\rm rot}  =  \hat{U} \hat{H}(t) \hat{U}^{\dagger} - i \hbar \hat{U} \frac{\partial}{\partial t} \hat{U}^\dagger \equiv \hat{H}(t=0)  - \frac{\omega}{2} \hat{\sigma}_z.
\end{equation}
 Here the unitary operator is given by $\hat{U}=e^{i \omega \hat{\sigma}_z t/2}$ for ${\bf m}(t)$ precessing counterclockwise. The transformed Hamiltonian $\hat{H}_{\rm rot}$ is time-independent in the frame rotating with the magnetization. The Zeeman term $ \omega\hat{\sigma}_z/2$, which emerges uniformly in the central region of devices in Fig.~\ref{fig:setup} and their N leads, will spin-split the bands of the N leads, thereby providing an intuitively appealing  rotating frame~\cite{Chen2009} picture of pumping. In this picture, the N leads in the rotating frame are labeled by ($p,\sigma$) [$p=L,R$ and $\sigma=\uparrow,\downarrow$] and they are biased by the electrochemical potential differences  $\mu_L^\downarrow - \mu_R^\uparrow = \omega$ and $\mu_R^\downarrow - \mu_L^\uparrow = \omega$. Thus, these leads behave as effective half-metallic ferromagnets which emit or absorb only one spin species. The counterpropagating dc currents of spin-polarized electrons flowing from lead $\mu_p^\downarrow$ to lead $\mu_{p^\prime}^\downarrow$, where electrons precess in the magnetic field of $\hat{H}(t=0)$ frozen at an angle $\theta$ with respect to the \mbox{$z$-axis} in order to enter into oppositely polarized lead, can be computed using NEGF,~\cite{Chen2009} transmission matrices, or the tunneling Hamiltonian approach.~\cite{Tserkovnyak2008}

However, the rotating frame approach cannot be applied to systems containing SOCs (or any other source of spin flips) because unitary transformation would generate time-dependent SOC terms in Eq.~(\ref{eq:hrot}). Nevertheless, it serves as a useful tool to compare the range of validity of different pumping formulas because the transport problem defined by Hamiltonian Eq.~(\ref{eq:hamiltonian}) with $\gamma_{\rm RSO}=\lambda_{\rm ESO}=0$ can be solved exactly in the rotating frame. We start by first extending the NEGF-based formulas for pumped current in the rotating frame for devices with semi-infinite N leads treated in Ref.~\onlinecite{Chen2009} to those with semi-infinite F leads. This eventually makes it possible to understand the origin of the orders of magnitude discrepancy between predictions made in Ref.~\onlinecite{Xiao2008} and Ref.~\onlinecite{Chen2009} for the dc pumping voltages in the same type of MTJs.

Since the system in the rotating frame is stationary, NEGFs which depend there only on the time difference $t-t'$ can be Fourier transformed to work with functions of a single energy argument
\begin{eqnarray}
\textbf{G}^r_{\rm rot}(E) & = & [E \textbf{1} - \textbf{H}_{\rm rot} - {\bm  \Sigma}^r_{\rm rot}(E)]^{-1}, \\
\textbf{G}^<_{\rm rot}(E) & = &
\textbf{G}^r_{\rm rot}(E){\bm \Sigma}_{\rm rot}^<(E) \textbf{G}^a_{\rm rot}(E).
\end{eqnarray}
The retarded self energy in the rotating frame is obtained from the self-energy in the lab frame through a
simple shift of its argument
\begin{equation}
{\bm \Sigma}_{{\rm rot},p}^r(E)={\bm \Sigma}_p^r \left (E + \frac{1}{2} \omega \hat{\sigma}_z \right).
\end{equation}
The lesser self-energy in the rotating frame is then given by
\begin{equation}
{\bm \Sigma}_{{\rm rot},p}^<(E)=i f \left( E + \frac{1}{2} \omega \hat{\sigma}_z \right) {\bm \Gamma}_p \left(E + \frac{1}{2} \omega \hat{\sigma}_z \right).
\end{equation}
This leads to ${\bf G}^r_{\rm rot}(E) = {\bf G}^r (E + \omega \hat{\sigma}_z/2)$ for the retarded GF and $\textbf{G}^<_{\rm rot}(E)=i\sum_{p}
\textbf{G}^r_{\rm rot} \textbf{f} {\bm \Gamma}_{p} (E + \omega \hat{\sigma}_z/2) \textbf{G}^a_{\rm rot}$ for the lesser one, where
$\textbf{f}=f(E + \omega \hat{\sigma}_z/2)$ is the Fermi function in the rotating frame written as a $2 \times 2$ matrix in the spin space. Using ${\bf G}^r_{\rm rot}(E)$, ${\bf G}^<_{\rm rot}(E)$ and the following identity
\begin{eqnarray}
i\textbf{G}^r_{\rm rot}{\bm \Gamma}\left(E + \frac{1}{2}\omega\hat{\sigma}_z \right) \textbf{G}^a_{\rm rot} = \textbf{G}^r_{\rm rot} - \textbf{G}^a_{\rm rot},
\end{eqnarray}
leads to an expression for pumped charge current in lead $p$
\begin{eqnarray} \label{eq:currentrot}
I_{p}^{\rm RF} & = & e \int\limits_{-\infty}^{+\infty} dE\, \frac{f_{\downarrow}-f_{\uparrow}} {2} {\rm Tr} \, [{\bm \Gamma}_p \hat{\sigma}_{z} \textbf{G}_{\rm rot}^r(E) {\bm \Gamma} \textbf{G}^a_{\rm rot}(E) \nonumber \\
 \mbox{}&& - {\bm \Gamma}_p \textbf{G}^r_{\rm rot}(E){\bm \Gamma} \hat{\sigma}_{z} \textbf{G}^a_{\rm rot}(E)],
\end{eqnarray}
where ${\bm \Gamma} = \sum_p {\bm \Gamma}_p$. Here $f_{\uparrow} = f(E + \omega/2)$ and $f_{\downarrow}=f(E - \omega/2)$ are the diagonal elements of ${\bf f}(E)$.

Thus, according to Eq.~(\ref{eq:currentrot}) only electrons whose energies fall into the interval $[E_F-\omega,E_F+\omega]$ participate in pumping (at zero temperature). Because of this, our more general solution Eq.~(\ref{eq:central}), truncated $N_{\rm ph}=1$ to take into account only zero or single microwave photon exchange processes, gives identical result to Eq.~(\ref{eq:currentrot}) in the rotating frame approach assuming absence of spin-flip processes.

Similarly to Sec.~\ref{sec:onephoton}, we can take into account that $\hbar \omega \ll E_F$ for microwave frequencies which yields pumped current in the adiabatic limit (where current is proportional to $\omega$):
\begin{equation}\label{eq:currentrotad}
I_p^{\rm RF} = \frac{e\omega}{2} {\rm Tr} \, \{ {\bm \Gamma}_p [\hat{\sigma}_z, \textbf{G}^r {\bm \Gamma}] \textbf{G}^a \}.
\end{equation}
Here all matrices are computed at $E_F$ in the laboratory frame after we neglect their frequency dependence in the rotating frame by invoking the  adiabatic condition $\omega \rightarrow 0$.

\subsection{Comparison with adiabatic scattering theory}\label{sec:compareas}

\begin{figure}
\includegraphics[scale=0.3,angle=0]{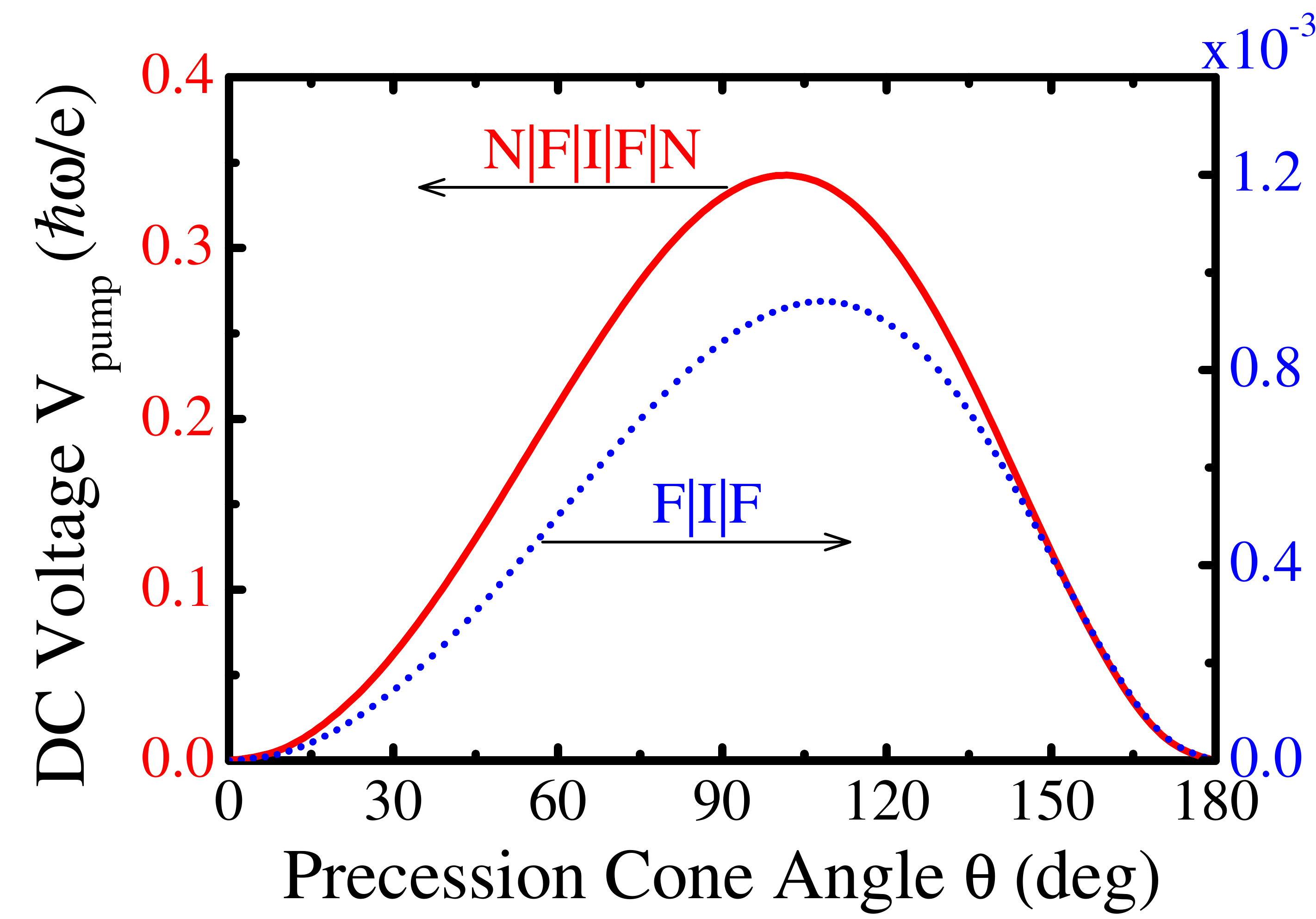}
\caption{(Color online) The comparison of the dc pumping voltage in a clean N$|$F$|$I$|$F$|$N junction with finite thickness F layers ($d_F=50$) and a clean F$|$I$|$F junction whose F layers are semi-infinite.~\cite{Xiao2008} The two curves can be computed using either the adiabatic NEGF formula in the rotating frame Eq.~(\ref{eq:currentrotad}) or the adiabatic scattering formula Eq.~(\ref{eq:brouwernegf}). The parameters of these junction are chosen as $E_F=-2\gamma$, $\Delta=2\gamma$, and $U_b=9\gamma$.}
\label{fig:bauerprb}
\end{figure}

In the adiabatic limit, one can also employ the Brouwer scattering formula~\cite{Brouwer1998} which gives the following expression for pumped charge current in terms of the derivatives of the instantaneous scattering matrix of the device:
\begin{equation}\label{eq:brouwer}
I_p^{\rm ST}  = e \omega \int\limits_0^{\tau } dt \,  {\rm Tr} \, \left \{ \sum_{q} \textbf{S}_{pq}(E_F,t) i \frac{\partial}{\partial t} \textbf{S}_{pq}^\dagger(E_F,t) \right \}.
\end{equation}
We can recast Eq.~(\ref{eq:brouwer}) in terms of NEGFs for stationary transport (which depend on only one energy argument) by using the Fisher-Lee formula~\cite{Fisher1981} for the scattering matrix
\begin{equation}
\textbf{S}_{pq} (E_F,t) = - \mathbf{1} \delta_{pq} + i \sqrt{\bm \Gamma}_p \cdot \mathbf{G}^r_{p q}(E_F,t) \cdot \sqrt{\bm \Gamma}_q.
\end{equation}
 Here $\mathbf{G}^r_{p q}$ is the submatrix of ${\bf G}^r=[E- {\bf H}(t) - {\bm \Sigma}^r]^{-1}$ which connects edge monolayer of the device attached to lead $p$ to the edge monolayer attached to lead $q$. The pumped current is then expressed as~\cite{Wang2003,Arrachea2006}
\begin{eqnarray}\label{eq:brouwernegf}
 I_p^{\rm ST} = e \omega \int\limits_0^\tau dt\,  {\rm Tr} \, \left\{ {\bm \Gamma}_p \textbf{G}^r(E_F,t) i\frac{\partial}{\partial t} [{\bm \Gamma} \textbf{G}^a(E_F,t)] \right\}.
\end{eqnarray}
This expression is equivalent to Eq.~(\ref{eq:currentrotad}) due to the fact that frequency dependence of all NEGF quantities in the latter has been neglected.

Nevertheless, the application of Eq.~(\ref{eq:brouwer}) to clean MTJs with no spin-flip processes, where the scattering matrix was obtained directly by matching the wave functions across a simple model of  Fe$|$MgO$|$Fe junction, has predicted~\cite{Xiao2008} three orders of magnitude smaller pumping voltage than the rotating frame formula Eq.~(\ref{eq:currentrotad}) applied to the same junction.~\cite{Tserkovnyak2008,Chen2009}  Figure~\ref{fig:bauerprb} explains that the origin of this discrepancy is not the particular formalism employed, but the assumed MTJ setup which contains  semi-infinite F leads in Ref.~\onlinecite{Xiao2008} and finite thickness F layers in Ref.~\onlinecite{Tserkovnyak2008,Chen2009}. Although the exact ratio of the voltage signals in these two models depends on the chosen values of $E_F$ and $\Delta$ at fixed $U_b$, the voltage signal in F$|$I$|$F MTJ model always remains below the one in the N$|$F$|$I$|$F$|$N model.

Both of these models describe unrealistic MTJs---there is no any kind of spin accumulation in the F layers in the F$|$I$|$F model, where semi-infinite F leads simply serve to define the spin-dependent scattering states, or spin accumulation persists throughout the finite thickness F layer in the N$|$F$|$I$|$F$|$N model. At the same time, the prediction of Ref.~\onlinecite{Xiao2008} is far below experimental values~\cite{Cao2009} ($\sim 1$ nV predicted versus $\sim 1$ $\mu$V  measured at $\simeq 2$ GHz FMR frequency), which points out to the need to take into account additional ingredients~\cite{Chui2008} in the MTJ model. 

\subsection{Comparison with continued fractions solution to double-time-Fourier-transformed NEGF equations}\label{sec:comapreguo}

The double-time-Fourier-transformed NEGF equations in Sec.~\ref{sec:onephoton} have been solved before for spin and charge pumping problems in an iterative manner using continued fractions.~\cite{Wang2003,Li2007a,Hattori2008} Therefore, in this Section we discuss the advantage of our solution from Sec.~\ref{sec:onephoton} over continued fractions technique using the same F$|$I$|$F clean MTJs model (with F layers of finite thickness and no SOCs) from Fig.~\ref{fig:bauerprb} as a test bed.

\begin{figure}
\includegraphics[scale=0.3,angle=0]{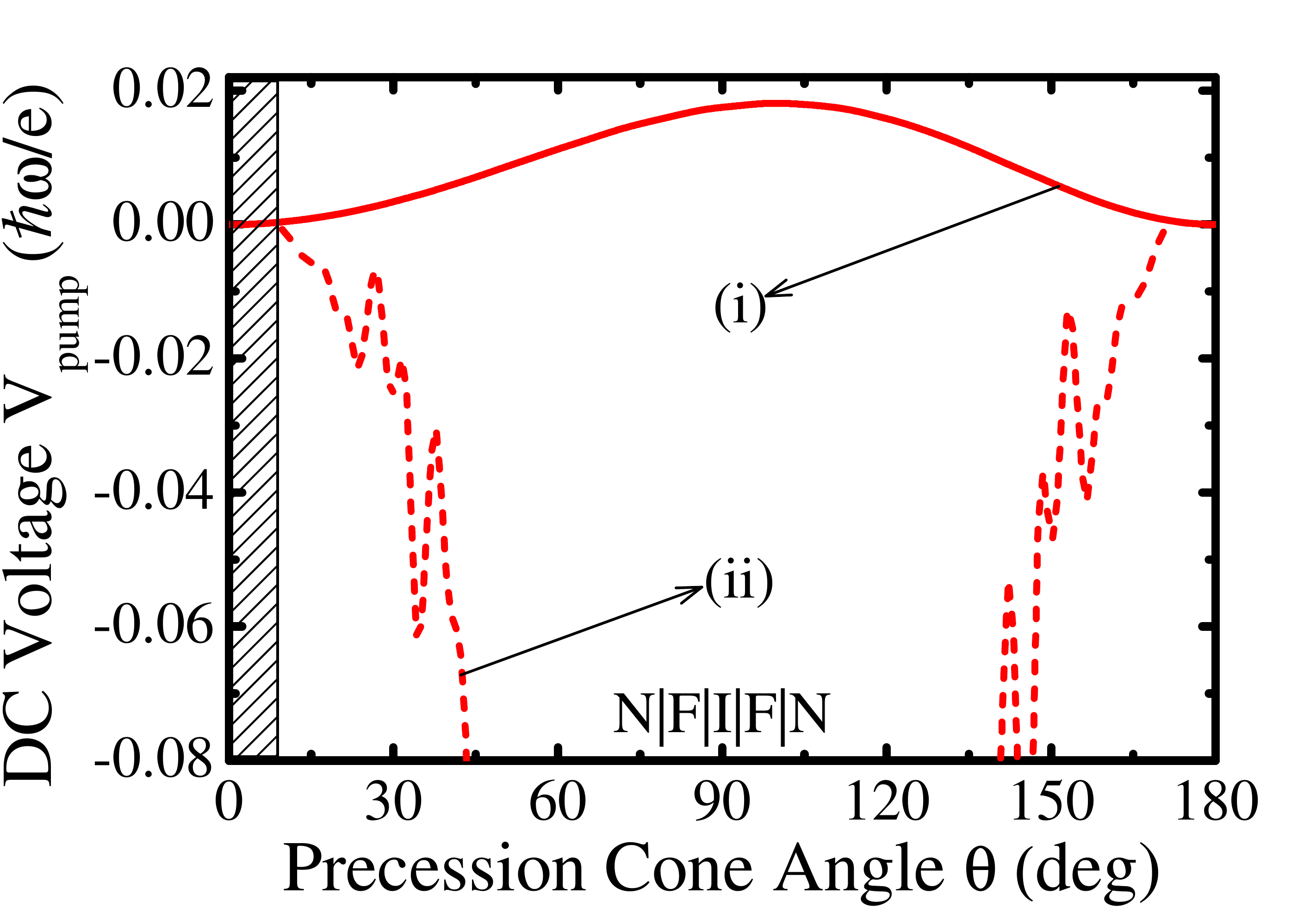}
\caption{(Color online) The dc pumping voltage in a clean F$|$I$|$F MTJ with finite thickness F layers ($d_F=50$) in the absence of any SOCs computed using: (i) the exact solution Eq.~(\ref{eq:currentrotad}) obtained via the rotating frame approach or, equivalently, full time-dependent solution Eq.~(\ref{eq:centraladiabatic}) with one photon processes taken into account $N_{\rm ph}=1$; and (ii) truncated (to $n=\pm 1$) continued fractions solution to double-time-Fourier-transformed NEGF equations which gives pumped charge current via Eq.~(\ref{eq:currentcfpert}). The shaded area marks the interval of precession cone angles $\theta \lesssim 10^\circ$ beyond which the continued fractions solution is not applicable anymore.}
\label{fig:cf}
\end{figure}

In the continued fractions method, one starts from the equation of motion for ${\bf G}^r_n(E)$ written as
\begin{eqnarray} \label{eq:ftdyson_wang}
{\bf G}^r_n(E) & = & 2\pi \delta (n\omega) {\bf g}^r_n(E) + {\bf G}^r_{n+1}(E){\bf V} {\bf g}^r_n(E) \nonumber \\
\mbox{} && + {\bf G}^r_{n-1}(E){\bf V}^\dagger {\bf g}^r_n (E),
\end{eqnarray}
Here ${\bf g}^r(E) = [E - {\bf H}_0 - {\bm \Sigma}^r(E)]^{-1}$ is the Fourier transform of the retarded GF ${\bf g}^r(t-t')$ in the absence of the pumping potential ${\bf H}'(t)=0$ which, therefore, depends only on the time difference $t-t'$ and can be Fourier transformed to a single energy argument. We also use notation ${\bf g}_n^r(E)={\bf g}^r(E+n\omega)$, ${\bf G}_n^r = 2\pi\delta(0)\bar{\bf G}^r_n$ and ${\bm \Sigma}^r_0 = {\bf V}^\dagger {\bf g}_1^r{\bm \alpha}_1 {\bf V} +  {\bf V} {\bf g}_{-1}^r {\bm \beta}_{-1} {\bf V}^\dagger$. This equation is then solved~\cite{Wang2003,Li2007a,Hattori2008} in an iterative manner using $\bar{\bf G}^r_0  =  [({\bf g}^r_0)^{-1} - {\bm \Sigma}^r_0]^{-1}$ for $n=0$, $\bar{\bf G}_n^r = \bar{\bf G}^r_{n-1} {\bf V}^\dagger {\bf g}_n^r {\bm \alpha}_n$ for $n \ge 1$, and $\bar{\bf G}_n^r = \bar{\bf G}^r_{n+1} {\bf V} {\bf g}_n^r {\bm \beta}_n$ for $n \le -1$. The coefficients ${\bm \alpha}_n$ and ${\bm \beta}_n$ are generated through continued fractions, ${\bm \alpha}_n ({\bf 1} - {\bf V}^\dagger {\bf g}^r_{n+1} {\bm \alpha}_{n+1} {\bf V} {\bf g}_n^r) = {\bf 1}$   and ${\bm \beta}_n ({\bf 1} - {\bf V} {\bf g}^r_{n-1} {\bm \beta}_{n-1} {\bf V}^\dagger {\bf g}_n^r) = {\bf 1}$.

The knowledge of $\bar{\bf G}^r_n$ allows one to express the pumped charge current in lead $p$ as~\cite{Wang2003,Hattori2008}
\begin{eqnarray}\label{eq:currentcf}
 I_p^{\rm CF}  & = &  \frac{1}{2\pi} \sum_{n=-\infty}^{+\infty} \int\limits_{-\infty}^{+\infty} dE\, {\rm Tr}\, \{ {\bm \Gamma}_p(E) \bar{\bf G}_n^r(E)  \nonumber \\
 \mbox{} && \times {\bm \Gamma}(E+n\omega) \bar{\bf G}_n^a(E) \} [f(E+n\omega) - f(E)].
 \end{eqnarray}
The summation over $n$ in this formula shows how multiphoton exchange processes assist current pumping. This expression can be used for non-adiabatic external potentials,~\cite{Wang2003} while in the adiabatic regime $\omega \rightarrow 0$ and at zero temperature the difference of Fermi functions is replaced by $f(E+n\omega) - f(E) \approx \omega  \delta(E-E_F)$, so that only the Fermi level states carry the pumped current.

Although one can in principle solve continued fractions for ${\bm \alpha}_n$ and ${\bm \beta}_n$ to arbitrary order $n$, this is virtually impossible to execute for sizable 3D devices (such as the ones in Fig.~\ref{fig:setup}) due to the need to compute numerous submatrices of ${\bf G}^r_{n-1}$ required to obtain ${\bf G}^r_n$. Instead, most of recent applications~\cite{Li2007a,Hattori2008} of the continued fractions solution to spin pumping in 2DEGs with the Rashba SOC have utilized only a few fractions ($|n| \le 3$). In other words, the convergence of the sum over $n$ in Eq.~(\ref{eq:currentcf}) can be achieved quickly only for small amplitude of the external potential $||{\bf V}|| \rightarrow 0$ which ensures that higher order fractions are negligible. The lowest order  $n = 0,\pm 1$ version of  Eq.~\eqref{eq:currentcf} simplifies to~\cite{Hattori2008}
\begin{eqnarray}\label{eq:currentcfpert}
 I_p^{\rm CF} & = & \frac{1}{2 \pi}  \int\limits_{-\infty}^{+\infty} dE\, {\rm Tr}\, \{ {\bm \Gamma}_p(E) {\bf g}^r(E) {\bf V}^\dagger {\bf A}_1(E) {\bf V}{\bf g}^a(E) \}
\nonumber \\
\mbox{}&& \times[f(E+\omega) - f(E)] \nonumber \\
\mbox{}&& + \frac{1}{2 \pi}  \int\limits_{-\infty}^{+\infty} dE\, {\rm Tr}\, \{ {\bm \Gamma}_p(E) {\bf g}^r(E) {\bf V} {\bf A}_{-1}(E) {\bf V}^\dagger {\bf g}^a(E) \} \nonumber \\
\mbox{}&& \times[f(E-\omega) - f(E)], \nonumber \\
\end{eqnarray}
where ${\bf A}_n(E) = {\bf g}^r_n(E) {\bm \Gamma}(E+n\omega) {\bf g}^a_n(E)$.

Figure~\ref{fig:cf} shows that Eq.~(\ref{eq:currentcfpert}) is insufficient to analyze pumping by magnetization dynamics in MTJs since it fails to reproduce the exact solution for dc pumping voltage in junctions with no spin flips given by Eq.~(\ref{eq:currentrot}) in the rotating frame approach. Because the strength of the pumping potential in Eq.~(\ref{eq:potential}) is determined by $\Delta \sin \theta$, pumping voltage computed from  Eq.~(\ref{eq:currentcfpert}) can be valid only at small angles cone angles ($\theta \lesssim 10^\circ$ in Fig.~\ref{fig:cf}; this interval would be somewhat larger in F$|$N multilayers). Even at small cone angles, the prediction $I_p^{\rm CF} \propto \Delta^2 \sin^2 \theta$ stemming from Eq.~(\ref{eq:currentcfpert}), which is in accord with the ``standard model''  Eq.~(\ref{eq:isin}), becomes incorrect in the presence of SOC where $I_p$ vs. $\theta$ turns out to be quite different (see Sec.~\ref{sec:rashba}). We note that one could try to use more general Eq.~\eqref{eq:currentcf}, but this would require to compute continued fractions ${\bm \alpha}_n$ and ${\bm \beta}_n$ to high order $n$, unlike our {\em non-perturbative} solution Eq.~\eqref{eq:central} which reproduces the exact result in the rotating frame using only $n=0,\pm1$ in the multiphoton GF in Eq.~\eqref{eq:multiphotongf}.

\section{The effect of interfacial Rashba SOC on the voltage signal of spin pumping in F$|$I$|$N and F$|$I$|$F junctions} \label{sec:rashba}

Equation~(\ref{eq:centraladiabatic}) applied to clean F$|$I$|$N junctions with interfacial Rashba SOC allows us to understand how the dc pumping voltage can appear in such semi-MTJs at the adiabatic level. The Rashba SOC is present at the F$|$I interface (i.e., at the last monolayer of the precessing F layer that is in contact with the tunnel barrier I) and intrinsically participates in the pumping process. This is in contrast to other recent theories~\cite{Takeuchi2010} of spin pumping in F$|$N multilayers where SOC is located away from the precessing F layer and, therefore, is not essential to understand the pumping effect itself.
 
\begin{figure}
\includegraphics[scale=0.34,angle=0]{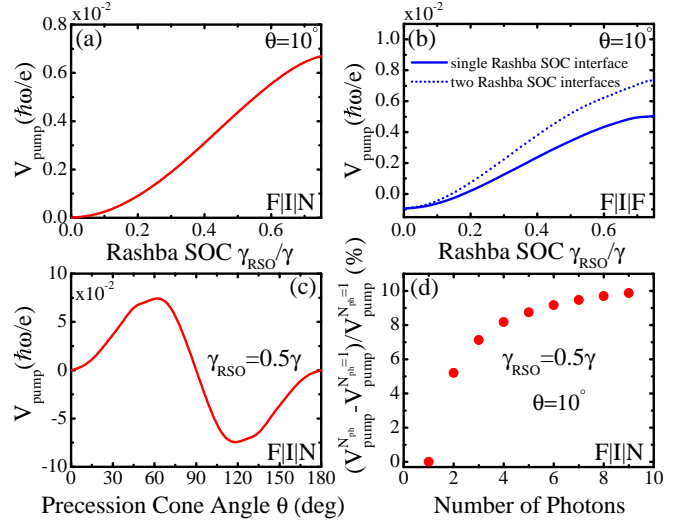}
\caption{(Color online) The dc pumping voltage in clean F$|$I$|$N semi-MTJ [panels (a),(c) and (d) and F$|$I$|$F MTJ [panel (b)] with finite thickness ($d_F=50$) F layers and non-zero interfacial Rashba SOC. The Rashba SOC is located within the last monolayer of the precessing F layer that is in contact with the tunnel barrier I in (a),(c),(d) [as illustrated in Fig.~\ref{fig:setup}(a)], or such edge monolayers are present in the left or both F layers [as illustrated in Fig.~\ref{fig:setup}(b)]  in panel (b). The data in panels (a),(b),(c) is computed by considering only one microwave photon exchange processes, while in panel (d) we show correction to this result when up to 9 microwave photons are taken into account in Eq.~(\ref{eq:centraladiabatic}) applied to F$|$I$|$N semi-MTJ.}
\label{fig:rashba}
\end{figure}

Figure~\ref{fig:rashba}(a) demonstrates that dc pumping voltage $V_{\rm pump} \propto \hbar\omega$ in F$|$I$|$N junctions emerges as soon as the Rashba SOC is ``turned on''. This could explain signal observed experimentally~\cite{Moriyama2008} in F$|$I$|$N junctions, in contrast to previous attempts~\cite{Tserkovnyak2008,Chen2009} which have predicted $V_{\rm pump} \propto (\hbar\omega)^2$. For comparison, Fig.~\ref{fig:rashba}(b) shows how the presence of strong Rashba SOC directly at the F$|$I interface also enhances $V_{\rm pump}$ in conventional F$|$I$|$F MTJs.

Figure~\ref{fig:rashba}(d) provides additional insight into the charge pumping mechanism where we show that the dc pumping voltage in F$|$I$|$N semi-MTJs requires to include exchange of up to ten microwave photons in order to reach its asymptotic value. However, since that asymptotic value of $V_{\rm pump}$ is only about 10\% higher than the result plotted in Figs.~\ref{fig:rashba}(a) and ~\ref{fig:rashba}(c), where only zero or single microwave photon exchange processes are taken into account, we employ only this lowest order approximation in the rest of the paper since [$N_{\rm ph}=1$ in Eq.~(\ref{eq:centraladiabatic}) is computationally much less expensive than $N_{\rm ph}=10$].

The unique experimentally testable signature of charge pumping in F$|$I$|$N semi-MTJs that we predict in Fig.~\ref{fig:rashba}(c) is angular dependence of $V_{\rm pump}(\theta)$ which changes sign and it is, therefore, quite different from the usual $V_{\rm pump}(\theta)$ for F$|$I$|$F MTJs shown in Fig.~\ref{fig:bauerprb}. The charge pumping in semi-MTJs with weak interfacial Rashba SOC can be obtained analytically using the second-order perturbation expansion of GF entering Eq.~(\ref{eq:brouwer}) as the version of the Brouwer scattering formula:
\begin{eqnarray}\label{eq:fincurrent}
I_L & = & e\hbar\omega D^{L} \int_0^\tau dt\, {\bf e}_z \cdot {\bf m}(t) \left[{\bf m}(t) \times \frac{\partial {\bf m}(t)}{\partial t}\right] \\ \nonumber
& = &  e \hbar \omega D^{L} \left [ \cos^2 \chi - \frac{1}{2} \sin^2 \chi \right] \sin^2 \theta \cos \theta.
\end{eqnarray}
Here $\chi$ is the angle between the axis around which the magnetization precesses and the direction of transport [in the case of our device in Fig.~\ref{fig:setup}(b),  $\cos^2 \chi - \frac{1}{2} \sin^2 \chi = -1/2$]. The lengthy explicit expression for $D^L \propto \gamma_{\rm RSO}^2$ determining the magnitude of the pumped current is provided in Appendix~\ref{sec:appendix}. In Fig.~\ref{fig:rashba}(c), we assume strong interfacial Rashba SOC ($\gamma_{\rm RSO} =0.5 \gamma$) so that $V_{\rm pump}(\theta)$ vs. $\theta$ plotted there deviates from this analytical expression $V_{\rm pump}(\theta) \propto \sin^2 \theta \cos \theta/G(\theta)$. Note that for small TAMR [$\gamma_{\rm RSO} \lesssim 0.4$ according to Fig.~\ref{fig:tamr}(c)] $G(\theta)$ can be considered nearly constant, so that $V_{\rm pump}(\theta) \propto \sin^2 \theta \cos \theta$ according to Eq.~\eqref{eq:fincurrent}.

The second-order nature of this process can be illustrated using real space Feynman paths where electron impinging onto the tunnel barrier is reflected with rotation of its spin introduced by the Rashba interface. Therefore, it has to travel twice through this monolayer to reach the right N lead. This picture is encoded quantitatively in the expression for $D^{L}$ in Eq.~(\ref{eq:dl}) which contains $\propto \gamma_{\rm RSO}^2$ dependence. We note that the same $\propto \gamma_{\rm RSO}^2$ and angular dependence has also been predicted~\cite{Manchon2011} for linear-response STT in F$|$I$|$N semi-MTJs with the interfacial Rashba SOC, which is in accord with reciprocal nature of STT and spin pumping. That is, observation of one of these two effects implies, by Onsager reciprocal relations, the existence of the other effect.~\cite{Hals2010}

\section{Disorder and extrinsic SOC effects on charge pumping in magnetic tunnel junctions}\label{sec:disorder}

\begin{figure}
\includegraphics[scale=0.35,angle=0]{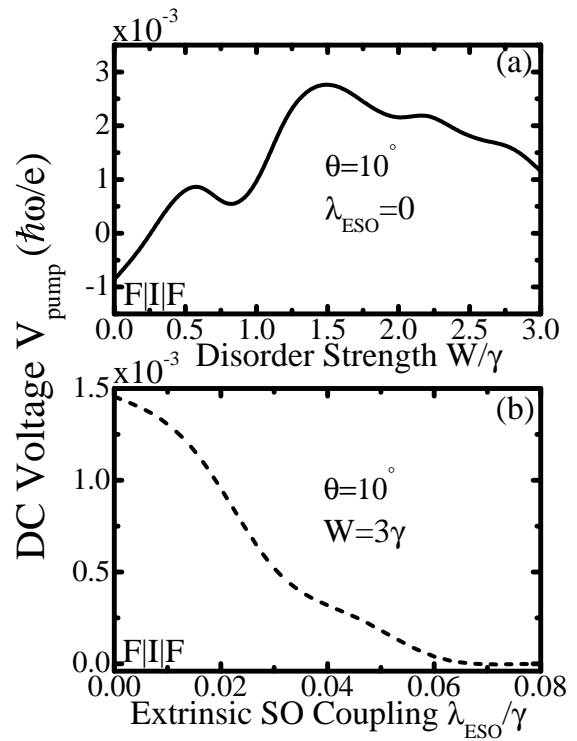}
\caption{The dc pumping voltage in F$|$I$|$F MTJs of finite thickness F layers ($d_F=50$) with: (a) static disorder of strength $W$ within F layers; (b) static disorder of strength $W=3\gamma$ ensuring diffusive transport regime (see Fig.~\ref{fig:fano}) and the extrinsic SOC of strength $\lambda_{\rm ESO}$ determined by such disorder via Eq.~(\ref{eq:eso}). The tunnel barrier I in both panels contains binary alloy disorder $\delta U_b = 0.5 \gamma$. The spin-diffusion length corresponding to the values of $\lambda_{\rm ESO}$ is shown in the inset of Fig.~\ref{fig:pz}.}
\label{fig:fig9}
\end{figure}

In this Section, we analyze how disorder and the corresponding extrinsic SOC affects dc pumping voltage in conventional F$|$I$|$F MTJs. To isolate their effects only, we assume that interfacial intrinsic Rashba SOC studied in Sec.~\ref{sec:rashba} is absent. When extrinsic SOC is negligible, we find that dc pumping voltage plotted in Fig.~\ref{fig:fig9}(a) is increasing in the quasiballistic transport regime (characterized by the Fano factor $F<1/3$ in Fig.~\ref{fig:fano}) and then decreases once the diffusive regime (characterized by the Fano factor $F=1/3$ in Fig.~\ref{fig:fano}) is reached. At first site, this initial increase of $V_{\rm pump}$ with increasing disorder is counterintuitive, even though conductance also decreases with disorder, since pumped current appears to be increasing with $W$. However, it can be explained qualitatively as being due to random electron scattering in real space which prolongs the average time an electron remains in the left F layer where it can interact with photons of the microwave pumping field. Similar enhancement of pure spin current pumping has been noticed in the diffusive regime in related device setups.~\cite{Hattori2008,Hattori2007}

The same disorder used in Fig.~\ref{fig:fig9}(a) is related to the extrinsic SOC through Eq.~(\ref{eq:eso}), which becomes a relevant effect if $\lambda_{\rm ESO}$ is renormalized by the band structure effects to become stronger than its vacuum value by several orders of magnitude.~\cite{Nagaosa2010} Unlike the interfacial Rashba SOC studied in Sec.~\ref{sec:rashba} which brings novel effects into the pumping mechanism, the extrinsic SOC simply reduces the dc pumping voltage in F$|$I$|$F junctions, as shown in Fig.~\ref{fig:fig9}(b). Our unified  quantum transport treatment of spin pumping [Fig.~\ref{fig:fig9}(b)] and spin diffusion (Fig.~\ref{fig:pz}) shows that voltage signal of spin pumping in MTJs is brought to negligible value when  the ratio of the F layer thickness to $L_{\rm sf}$ is $d_F/L_{\rm sf} \simeq 5$. We emphasize that our fully quantum-mechanical treatment of the conduction electrons is necessary to understand such interplay of spin pumping, spin accumulation around interfaces~\cite{Chui2008} and spin diffusion in MTJs since conventional approach~\cite{Tserkovnyak2005} developed for F$|$N multilayers, where pumping is treated quantum-mechanically while subsequent propagation of spins and charges is described semiclassically using phenomenological mean free path and spin-diffusion length, is inapplicable to systems containing tunnel barriers where spin accumulation is not well-defined.~\cite{Taniguchi2008a}

\section{Concluding Remarks} \label{sec:conclusions}

We have derived an exact and efficient for computational implementations solution to the equations of motion for the double-time-Fourier-transformed NEGFs in the presence of time-periodic external potential. Unlike continued fractions solution~\cite{Wang2003} for the same equations, which is often applied to problems of spin~\cite{Hattori2008} and charge pumping~\cite{Wang2003} by computing only  a finite number of continued fractions while assuming that the amplitude of time-periodic external potential is small, our formulas for pumped charge Eq.~(\ref{eq:central}) and spin Eq.~(\ref{eq:centralspin}) currents in the leads of a multiterminal devices  can be used for {\em arbitrary strength} of periodic driving potential (thereby covering both perturbative~\cite{Hattori2008,Li2007a} and non-perturbative regimes) or frequency (thereby covering both adiabatic and non-adiabatic pumping regimes).

This fully quantum-mechanical treatment of pumping processes and subsequent propagation of electrons is applied to the problem of charge pumping by precessing magnetization in the single F layer of F$|$I$|$N semi-MTJ or  F$|$I$|$F conventional MTJ in the presence of intrinsic Rashba SOC at the F$|$I interface. The non-zero interfacial Rashba SOC, located~\cite{Gmitra2009} within the edge monolayer of the precessing F in contact with the tunnel barrier I, generates non-zero dc pumping voltage in F$|$I$|$N semi-MTJ at the adiabatic level (i.e., pumping voltage is proportional to the microwave frequency $\omega$). This could explain observations of voltage signal with such properties in the recent experiments~\cite{Moriyama2008} on microwave-driven F$|$I$|$N semi-MTJs where previously formulated theories~\cite{Tserkovnyak2008,Chen2009} have found only a very small non-adiabatic ($\propto \omega^2$) voltage signal. We further predict a {\em unique signature} of this charge pumping phenomenon---the pumped charge current changes sign ($I \propto \sin^2 \theta \cos \theta$ for small $\gamma_{\rm RSO}$) as the function of the precession cone angle $\theta$ so that measuring the corresponding dc pumping voltage ($V_{\rm pump} \propto \sin^2 \theta \cos \theta$ for small $\gamma_{\rm RSO}$) would confirm our prediction.

Besides offering quantitative description of charge and spin pumping processes, our solution for NEGFs whose two energy arguments are connected by the Floquet theorem describing multiphoton emission and absorption processes also provides new physical insights: ({\em i}) in the absence of SOCs, emission or absorption of {\em one photon} is sufficient to match the exact solution in the rotating frame;~\cite{Chen2009,Hattori2007}   ({\em ii}) in the presence of Rashba SOC, exchange of up to ten photons is required to reach asymptotic value of the pumped currents. Nevertheless, this asymptotic value is only about 10\% larger than the value obtained using just one photon processes in the presence of SOCs.

We also find that static disorder can increase the dc pumping voltage in F$|$I$|$F MTJs with finite thickness F layers in the quasiballistic transport regime where scattered electrons spend more time within the precessing F layer to interact with microwave photons. The extrinsic SOC determined by the impurity potential responsible for the diffusive transport regime causes spin relaxation which ultimately diminishes the pumping voltage in F$|$I$|$F MTJs to zero when the spin-diffusion length is about five times shorter than the thickness of the F layers.

\begin{acknowledgments}
We thank G.~E.~W.~Bauer, S.-H. Chen, T.~Moriyama, M. Wimmer and J. Q. Xiao for illuminating discussions. F. M. and B. K. N. were supported by DOE Grant No. DE-FG02-07ER46374. J. F. was supported by Grant DFG SFB 689.  N. N. was supported by Grant-in-Aids for Scientific Research (21244053) from the Ministry of Education, Culture, Sports, Science and Technology of Japan, Strategic International Cooperative Program (Joint Research Type) from Japan Science and Technology Agency, and also by Funding Program for World-Leading Innovative R\&D on Science and Technology (FIRST Program).
\end{acknowledgments}

\appendix

\section{Expression for pumped charge current in F$|$I$|$N junctions with interfacial Rashba SOC}\label{sec:appendix}

Here we provide explicit expression for the coefficient $D^{L}$ which determines the strength of pumped charge current in  Eq.~(\ref{eq:fincurrent}) for F$|$I$|$N semi-MTJ with weak interfacial Rashba SOC, as discussed in Sec.~\ref{sec:rashba}. In the
absence of Rashba SOC, the retarded GF of F$|$I$|$N semi-MTJ can be written as
\begin{equation}\label{eq:rgf}
{\bf G}^r_0 = {\bf g}_0 + {\bf g}_1 {\bf m} \cdot \hat{\bm \sigma}.
\end{equation}
Starting from this expression, the second-order perturbation theory in the powers of $\gamma_{\rm RSO}$ applied to GF in the Brouwer scattering formula, recast as Eq.~(\ref{eq:brouwer}), yields
\begin{eqnarray}\label{eq:dl}
D^{L} & = & 4\sum_\alpha {\rm Re}\, \{ {\rm Tr}\, [{\bm \Gamma}_p \textbf{g}_0 V_\alpha \textbf{g}_1 V_\alpha \textbf{g}_1{\bm \Gamma} \textbf{g}^\dagger_1 \nonumber \\
&& -{\bm \Gamma}_p \textbf{g}_1 V_\alpha \textbf{g}_1 V_\alpha \textbf{g}_0 {\bm \Gamma} \textbf{g}^\dagger_1 \nonumber
 -{\bm \Gamma}_p \textbf{g}_1 V_\alpha \textbf{g}_1 {\bm \Gamma} \textbf{g}^\dagger_0 V_\alpha \textbf{g}^\dagger_1  \nonumber \\
&& -{\bf\Gamma}^{\alpha}\textbf{g}_1 V_\alpha \textbf{g}_1 {\bm \Gamma} \textbf{g}^\dagger_1 V_\alpha \textbf{g}^\dagger_0] \},
\end{eqnarray}
Here the vector \mbox{${\bf V}=(V_x,V_y,V_z)$}, which contains the strength of the Rashba SOC $\gamma_{\rm RSO}$, is defined by \mbox{$\hat{H}_{\rm RSO} = {\bf e}_x \cdot ({\bf V} \times \hat{\bm \sigma})$} where  $\hat{H}_{\rm RSO}$ is the Rashba
Hamiltonian [i.e., the fourth term in Eq.~(\ref{eq:hamiltonian})].





\end{document}